\documentclass[aps,prx,twocolumn,superscriptaddress]{revtex4-2}
\usepackage{amsmath}
\usepackage{amssymb}
\usepackage{amsthm}
\usepackage{amsfonts}
\usepackage{listings}
\usepackage{physics}
\usepackage{enumerate}
\usepackage{latexsym}
\usepackage{newunicodechar}
\newunicodechar{⋅}{\cdot}
\usepackage{color}
\usepackage{bm}
\usepackage{hyperref}

\newcounter{suppfigure}

\hypersetup{
 pdfnewwindow=true, colorlinks=true,
 linkcolor=blue, anchorcolor=blue,
 citecolor=blue, filecolor=blue,
 menucolor=blue, urlcolor=blue}

\usepackage{psfrag}

\usepackage{bm}
\usepackage{graphicx}
\usepackage{subfigure}

\usepackage{multirow}
\usepackage{array}

\RequirePackage[normalem]{ulem} 
\RequirePackage{color}\definecolor{RED}{rgb}{1,0,0}\definecolor{BLUE}{rgb}{0,0,1}

\def\eg{{\it e.g.}\ }

\input{epsf}

\newcommand{\nc}{\newcommand}
\nc{\webirvsp}{\href{https://github.com/zjwang11/irvsp}{\texttt{IRVSP}} }
\nc{\webirtb}{\href{https://github.com/zjwang11/irvsp}{\texttt{ir2tb}} }
\nc{\webirpw}{\href{https://github.com/zjwang11/ir2pw}{\texttt{ir2ph}} }
\nc{\webchecktopmat}{\href{https://www.cryst.ehu.es/cryst/checktopologicalmagmat}{\texttt{Check Topological Mat}}}
\nc{\webposabr}{\href{https://github.com/zjwang11/UnconvMat/blob/master/src_pos2aBR.tar.gz}{\texttt{POS2ABR}} }
\nc{\webUnconvMat}{\href{http://tm.iphy.ac.cn/UnconvMat.html}{\texttt{UnconvMat}} }
\nc{\online}{\href{http://tm.iphy.ac.cn/UnconvMat.html}{online}}

\begin{document}

\tolerance 10000


\title{Spin-Polarized Weyl Fermions and Unconventional Spin Transport}
\author{Chenhao~Liang}
\affiliation{Beijing National Laboratory for Condensed Matter Physics and Institute of Physics, Chinese Academy of Sciences, Beijing 100190, China}
\affiliation{School of Physical Sciences, University of Chinese Academy of Sciences, Beijing 100049, China}

\author{Haohao~Sheng}
\email{hhsheng@iphy.ac.cn}
\affiliation{Beijing National Laboratory for Condensed Matter Physics and Institute of Physics, Chinese Academy of Sciences, Beijing 100190, China}

\author{Sheng Zhang}
\affiliation{Beijing National Laboratory for Condensed Matter Physics and Institute of Physics, Chinese Academy of Sciences, Beijing 100190, China}
\affiliation{School of Physical Sciences, University of Chinese Academy of Sciences, Beijing 100049, China}
 
\author{Zhong Fang}
\affiliation{Beijing National Laboratory for Condensed Matter Physics and Institute of Physics, Chinese Academy of Sciences, Beijing 100190, China}
\affiliation{School of Physical Sciences, University of Chinese Academy of Sciences, Beijing 100049, China}

\author{Hongming Weng}
\email{hmweng@iphy.ac.cn}
\affiliation{Beijing National Laboratory for Condensed Matter Physics and Institute of Physics, Chinese Academy of Sciences, Beijing 100190, China}
\affiliation{Condensed Matter Physics Data Center, Chinese Academy of Sciences, Beijing 100190, China}

\author{Zhijun~Wang}
\email{wzj@iphy.ac.cn}
\affiliation{Beijing National Laboratory for Condensed Matter Physics and Institute of Physics, Chinese Academy of Sciences, Beijing 100190, China}
\affiliation{Condensed Matter Physics Data Center, Chinese Academy of Sciences, Beijing 100190, China}

\date{\today}

\begin{abstract}
Large and electrically controllable spin splitting in nonmagnetic materials is highly desirable for condensed-matter physics and spintronics, enabling strongly spin-polarized quasiparticles and unconventional spin transport as well as their electrical control. Here, we establish a general framework for critical spin-polarized Weyl fermions in two-dimensional (2D) nonmagnetic systems. We predict the $\alpha$-Bi-like Group-V puckered-layer family of 2D ferroelectrics as a material platform combining critical spin-Weyl fermions, giant spin splitting, and electrically switchable nonlinear and spin transport. Our calculations reveal exchange-scale spin splitting induced by spin-orbit coupling and a $p$-wave spin texture in the BiAs monolayer. Moreover, BiAs hosts four critical spin-Weyl points, each formed within a single-spin channel. The nearly quantized $\langle S_z\rangle$ is protected by an emergent $S_z$-conservation symmetry in the low-energy limit. The spin-Weyl cones also give rise to unconventional spin-transport phenomena, including a purely out-of-plane Edelstein response and a pronounced spin Nernst effect. Ferroelectric reversal further enables electrical switching of the $p$-wave spin texture and inversion-odd responses, while the tunability of these properties through composition, strain, and magnetic field is also systematically demonstrated across the entire family. These results establish both a general nonmagnetic framework and a promising material platform for exploring critical spin-Weyl physics, nonlinear Hall phenomena, and unconventional spin transport.

\end{abstract}

\maketitle

\paragraph*{Introduction.}
Momentum-dependent spin splitting is a fundamental phenomenon in condensed-matter physics and provides the microscopic basis for spintronic functionalities~\cite{spintronics-doi:10.1126/science.1065389,spintronics-RevModPhys.76.323,spintronics-Awschalom2007}. It conventionally originates from either exchange fields in magnetically ordered systems~\cite{FMspin-PhysRev.49.537,FMspin-Stoner1938,liuSpontaneousFerromagnetismInduced2022,yaoAxionInsulatorWeyl2025} or relativistic spin-orbit coupling (SOC) in noncentrosymmetric crystals~\cite{socspin-PhysRev.100.580,socspin-Bihlmayer2022,shaoLargeSpinHall2023}. Exchange interactions generally produce substantially larger spin splittings than SOC, but magnetic materials inevitably involve magnetic domains, stray fields, and often limited electrical tunability~\cite{Schaefer2021Domains,NewHorizonsSpintronics2022,Matsukura2015NNano}. Realizing exchange-scale spin splitting and magnet-like spin transport in nonmagnetic materials therefore represents an important opportunity for spintronics.

Beyond its magnitude, spin splitting can endow the electronic structure with nontrivial band topology in momentum space. Of particular interest are fully spin-polarized Weyl fermions, where the crossed bands are in a single-spin channel~\cite{Liu2024PRB,Li2N-2025-vqdz-12bz}. In such states, the low-energy quasiparticles inherit a Weyl-like linear dispersion, while the spin polarization remains strongly locked to one spin direction rather than forming a conventional winding spin texture~\cite{Bychkov1984JETPL,Hsieh2009Science}. 
Fully spin-polarized Weyl fermions can naturally emerge in collinear magnetic systems described by spin space groups (SSGs). This raises an intriguing question: Can fully spin-polarized Weyl nodes also exist in nonmagnetic systems?
The spin-Weyl cones can host concentrated spin-resolved Berry curvature, large orbital magnetic moments, and unconventional magnetic, optical, and transport responses~\cite{Xiao2007PRL}.

In this work, we establish the symmetry principle and general classification of critical spin-Weyl fermions in two-dimensional (2D) nonmagnetic systems. 
Through systematic first-principles calculations, we establish the $\alpha$-Bi-like Group-V puckered-layer family of 2D ferroelectrics as a real material platform for critical spin-Weyl physics, large spin splitting, and electrically switchable nonlinear and spin transport. 
Crystal symmetry enforces robust out-of-plane spin-valley locking with a $p$-wave pattern.
The spin splitting is exceptionally large, reaching an exchange-scale value of $\sim0.6$ eV in BiAs. Moreover, BiAs is at a topological phase boundary with four critical spin-Weyl cones.
The tiny mass gaps of the Weyl cones generate giant nonlinear charge and spin Hall responses. 
The spin-Weyl cones with the $p$-wave pattern also produce a purely out-of-plane Edelstein response and a pronounced spin Nernst effect.
The tunability of these properties through composition, strain, magnetic field, and ferroelectric (FE) polarization is also systematically investigated across the whole family.

\begin{figure*}[!t]
	\centering
	\includegraphics[width=0.95\textwidth]{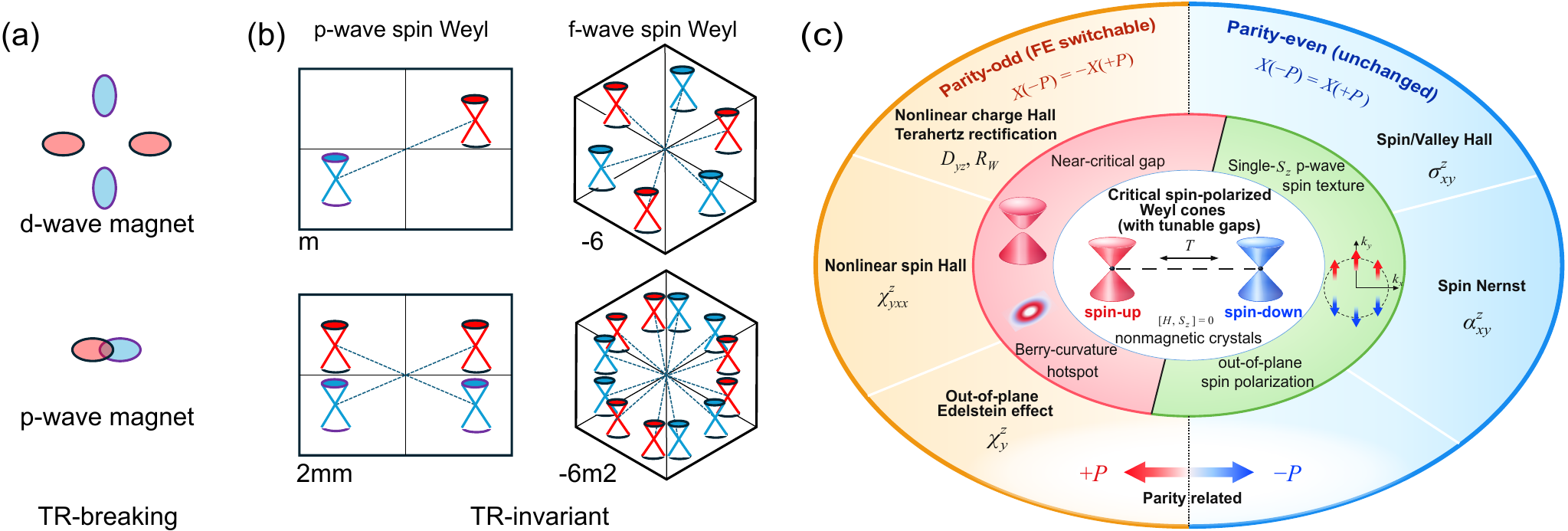}
\caption{
Odd-wave spin textures and spin-Weyl fermions in nonmagnetic systems. 
(a) Unconventional spin splitting in time-reversal (TR) breaking magnets. 
(b) $P$-wave and $f$-wave $S_z$-polarized Weyl cones in TR-invariant nonmagnetic materials. Red and blue denote opposite $S_z$ polarizations. (c) Spin-Weyl physics and applications in nonmagnetic materials. The spin-Weyl cones feature a single-spin band crossing and an odd-wave spin texture, giving rise to diverse charge and spin transport.
}\label{odd_wave}
\end{figure*}

\paragraph*{$k\cdot p$ models for critical spin-Weyl fermions.}
A related momentum-space structure has recently emerged in unconventional magnets~\cite{unConvenmag-PhysRevX.12.011028,unConvenmag-PhysRevX.12.031042,unConvenmag-PhysRevX.12.040501,ssg-szy-song2025unifiedsymmetryclassificationmagnetic,irssg}. In coplanar $p$-wave antiferromagnets (AFMs), SSG symmetries can produce odd-in-momentum spin splitting with collinear spin polarization constrained  perpendicular to the magnetic plane, yielding $S_z$-polarized $p$-wave Fermi surfaces with nonquantized $\langle S_z\rangle$ in the absence of SOC~\cite{pwave-hellenes2024pwavemagnets,pwave-Yamada2025,pwave-Chakraborty2025,pwave-PhysRevB.111.L161301}. Hyperspin altermagnets are further introduced with quantized hyperspin $\langle \hat{\mathcal S}_z\rangle$ in the absence of SOC~\cite{maHyperspinAltermagnets2026}. However, odd-wave (\eg, $p$- and $f$-wave) textures are compatible with time-reversal symmetry in principle [Figs.~\ref{odd_wave}(a,b)], allowing their realization without magnetic order. Such a nonmagnetic realization offers two advantages over magnets. First, in $p$-wave/hyperspin altermagnets, the $\langle S_z\rangle$ orientation is protected by a SSG symmetry, whereas SOC generally destroys this protection and allows finite in-plane components. By contrast, in 2D nonmagnetic crystals with an out-of-plane mirror symmetry $M_z$, only the $\langle S_z\rangle$ component is allowed even in the presence of SOC. 
Second, reversing the AFM order is generally difficult and can be further complicated by magnetic domains. Nonmagnetic FE materials give the opportunity to switch the spin texture by changing FE polarization. These advantages motivate the exploration of nonmagnetic spin-Weyl fermions and routes to electrical control.

Here, we derive the symmetry conditions for critical spin-Weyl fermions in 2D nonmagnetic systems.
When the low-energy valence and conduction orbitals share the same $M_z$ eigenvalue, the mirror representation reduces to $D(M_z)=-i\tau_0\sigma_z$.
The constraint $[H^{\mathrm{eff}}(\mathbf{k}),D(M_z)]=0$ forbids $\sigma_{x,y}$ terms, yielding $[H^{\mathrm{eff}}(\mathbf{k}),\hat{S}_z]=0$ and quantized $\langle S_z\rangle$ in the low-energy limit.
Time-reversal symmetry relates the two spin blocks, giving
$\mathcal{H}(\mathbf{k},\lambda)=\operatorname{diag}[h(\mathbf{k},\lambda),h^T(-\mathbf{k},\lambda)]$,
where $h(\mathbf{k},\lambda)$ is a $2\times2$ Hamiltonian in the spin-up sector. Here, $\lambda$ parametrizes strain, composition, or structural variation.
Then, one can obtain fully spin-polarized Weyl nodes in a general three-dimensional (3D) parameter space $(k_x,k_y,\lambda)$, which correspond to critical Weyl points in the 2D system. Unlike the massless Dirac/Weyl points in graphene~\cite{Novoselov2005Nature}, the critical Weyl points can readily acquire finite mass through strain engineering or chemical composition tuning~\cite{luRealizationTwodimensionalWeyl2024a}. Within this framework, we reveal that the spin-Weyl fermions are allowed in four point groups: $m$, $2mm$, $\bar{6}$, and $\bar{6}m2$ [Fig.~\ref{odd_wave}(b)].
The first two yield $p$-wave $S_z$-polarized spin textures, whereas the latter two yield $f$-wave patterns. We focus on $2mm$ in the main text [see more in Section~\textcolor{blue}{A} of the Supplementary Material (SM)].
Using the method of invariants~\cite{gaoIrvspObtainIrreducible2021,zhangZhangVASP2KPKpModels2023}, we construct the four-band $k\cdot p$ Hamiltonian as
\begin{equation}
\begin{aligned}
\mathcal{H}(\mathbf{k},\lambda)&={}
[\epsilon(\mathbf{k})\tau_0
+M(\mathbf{k})\tau_z
+A(\mathbf{k})\tau_x+\lambda k_x\tau_y]\sigma_0\\
&
+[b_1k_y\tau_0
+b_3k_y\tau_x+b_2k_y\tau_z+c_7k_xk_y\tau_y]\sigma_z,
\end{aligned}
\label{eq:kpmain}
\end{equation}
with $\epsilon(\mathbf{k})=a_1+c_1k_x^2+c_2k_y^2$,
$A(\mathbf{k})=a_2+c_3k_x^2+c_4k_y^2$ and $M(\mathbf{k})=a_3+c_5 k_x^2+c_6k_y^2$. The matrices $\tau$ and $\sigma$ act on the orbital and spin degrees of freedom, respectively. 

\paragraph*{Material~realization—BiAs~monolayer.}
\begin{figure}[!t]
	\centering
    \includegraphics[width=0.5\textwidth]{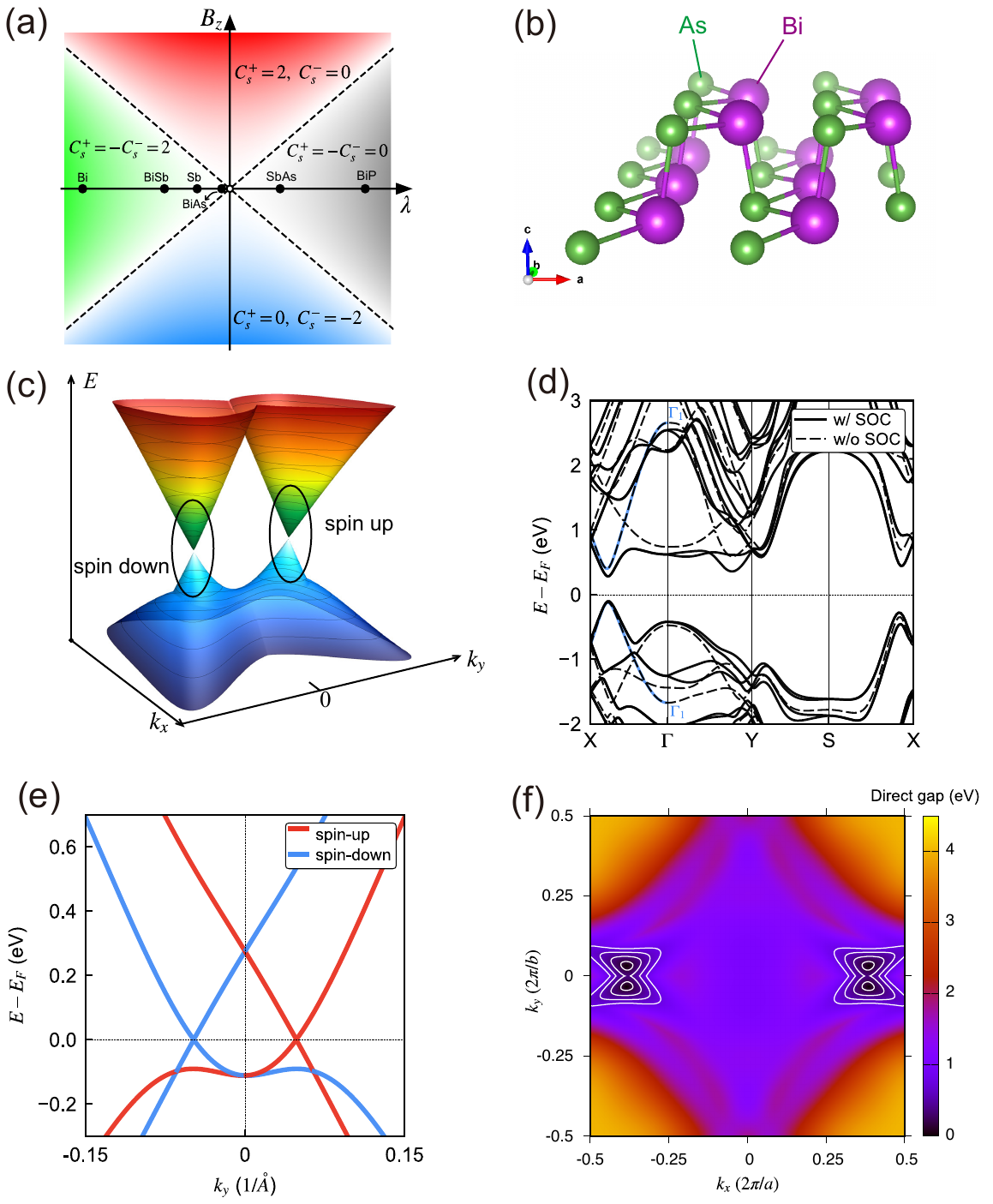}
\caption{Topological phases of the $\alpha$-Bi-like Group-V puckered-layer family, crystal structure, electronic spectra and band gaps of BiAs monolayer.
(a) Topological phase diagram as a function of the parameter $\lambda$ (\eg, strain, composition, or structural variation) and the out-of-plane magnetic field $B_z$. $C_{s}^+$ and $C_{s}^-$ indicate the partial spin Chern numbers, the spin Chern number is $C_s=(C_{s_z}^+ - C_{s_z}^- )/2$; the color intensity represents the band gap. 
(b) Puckered BP-like structure of BiAs.
(c) 3D view of the energy spectrum
around a pair of critical spin-Weyl nodes. 
(d) Electronic band structures along high-symmetry lines without (w/o) and with (w/) SOC.
The band-edge states on X-$\Gamma$ connect to the bands of the $\Gamma_1$ irreducible representation at $\Gamma$ in the absence of SOC.
(e) Band dispersions through a pair of critical spin-Weyl nodes ($k_x=k_x^W$). 
(f) Gap distribution in the first Brillouin zone with constant-energy contours (white lines). Four critical spin-Weyl points are located at $(\pm W_x, \pm W_y)=\left(\pm0.384, \pm0.031\right)$. 
}
\label{BiAs_band}
\end{figure}

The BiAs monolayer crystallizes in a puckered black-phosphorus-like (BP-like) structure with space group $Pmn2_1$ (\#SG~31; $\alpha$-Bi phase~\cite{zhaoRealization2DMetals2025a}), as shown in Fig.~\ref{BiAs_band}(b). This structure possesses a mirror $M_y$ and a screw rotation $C_{2x}$ (glide mirror $M_z$) followed by a fractional translation $(\frac{1}{2},\frac{1}{2})$ with respect to the lattice vectors.
The band structure on the high-symmetry lines in Fig.~\ref{BiAs_band}(d) appears to exhibit a sizable direct gap even in the presence of SOC. 
However, detailed calculations reveal four critical Weyl nodes away from these high-symmetry lines.
Their locations are clearly shown in the energy gap plot of Fig.~\ref{BiAs_band}(f).
The tiny gap ($\sim$ 3 meV) of the critical Weyl node indicates that the system lies at a topological phase boundary in Fig.~\ref{BiAs_band}(a).
By modifying $\lambda$ in Eq.~(\ref{eq:kpmain}), the mass term of the 2D critical Weyl node can be easily tuned across the members of the $\alpha$-Bi-like Group-V puckered-layer family.
The calculation methods are provided in Section~\textcolor{blue}{B} of the SM.   

\begin{figure}[!t]
	\centering
    \includegraphics[width=0.48\textwidth]{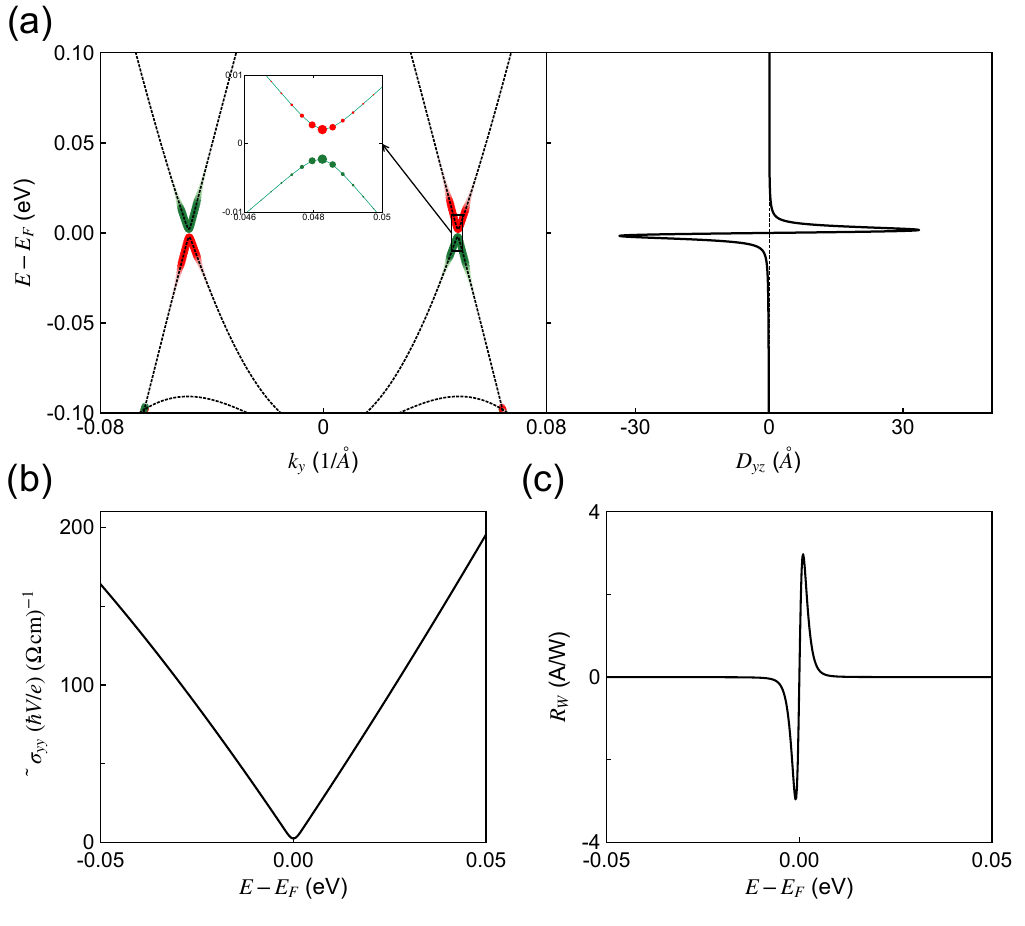}
\caption{Nonlinear charge Hall response and terahertz rectification.
(a) Weyl bands and the corresponding BCD $D_{yz}$ in BiAs. 
The circle size and color scale represent the magnitude and sign of the Berry curvature $\Omega_z$ on the bands.
The left inset shows a magnified view of the Weyl cone at $k_y=0.046~\mathrm{\AA}^{-1}$.
The BCD is calculated using \(k_BT=1~{\rm meV}\) (\(T=11.6~{\rm K}\)).
(b) Drude weight $\tilde{\sigma}_{yy}$ and (c) terahertz responsivity $R_W$ of intrinsic BiAs. 
A device width of $W=1~\mu{\rm m}$ is used.
}\label{BCD_strain}
\end{figure}

We then plot the band dispersions of Weyl nodes along the $k_y$ direction, as presented in Fig.~\ref{BiAs_band}(e). The Fermi velocity of the Weyl cones is about $9.0\times10^7$ ($5.6\times10^7$) cm/s along the $y$ ($x$) direction, comparable to that of graphene. 
The corresponding 2D dispersions are shown in Fig.~\ref{BiAs_band}(c).
The Weyl bands exhibit three key features.
First, due to the presence of $M_z$ symmetry, only the out-of-plane spin component $\langle S_z\rangle$ is allowed. 
Notably, the two bands forming each Weyl node carry the same $\langle S_z\rangle$, with a magnitude of about $0.9~\hbar/2$, indicating nonmagnetic spin-Weyl fermions within a nearly single-spin channel.
Second, even though the band spin splitting stems from SOC, the magnitude $\Delta_{\mathrm{SOC}}$ is remarkably large, reaching $\sim 0.6$ eV, as shown in Fig.~\ref{BiAs_band}(e). Third, spin-valley locking is clearly observed. The spin texture exhibits a $p$-wave pattern, implying an even more robust route (in electronic materials with SOC) to the unconventional spin transport associated with $p$-wave magnets. 
We further examine other members of the $\alpha$-Bi-like Group-V puckered-layer family.
They are dynamically stable, with no imaginary phonon modes [Section~~\textcolor{blue}{C} of the SM], and exhibit similar $S_z$-valley-locked band structures.
Across this family, the spin-Weyl gap evolves from positive (trivial $C_s=0$: SbAs, BiP, SiTe, and GeTe), through nearly zero (critical Weyl phase: BiAs), to negative (topological $C_s=2$: BiSb, $\alpha$-Sb, and $\alpha$-Bi), as presented in Fig.~\ref{BiAs_band}(a).
A negative gap denotes band inversion, resulting in a doubled quantum spin Hall (DQSH) phase with $C_s=2$~\cite{wangHourglassFermions2016}. The topological gaps reach $-195$, $-89$, and $-450$ meV for BiSb, $\alpha$-Sb, and $\alpha$-Bi monolayers [Section~~\textcolor{blue}{D} of the SM]. The spin-Weyl cones with giant spin splitting and unconventional $p$-wave spin textures can give rise to unconventional charge and spin transport in nonmagnetic systems.

\paragraph*{Berry curvature dipole and terahertz rectification.} 
The Weyl cones with tiny gaps generate pronounced Berry-curvature hotspots and induce a large  Berry curvature dipole (BCD), which provides the band-geometric origin of the nonlinear charge Hall effect. In BiAs, the $2mm$ point symmetry allows only the BCD in-plane component $D_{yz}$.
Fig.~\ref{BCD_strain}(a) shows the band-resolved Berry curvature and the corresponding $D_{yz}$. 
The Berry-curvature hotspots are concentrated in the low-energy Weyl nodes, leading to a sharp enhancement of the intrinsic BCD as the energy approaches the valley energy. The peak value reaches $D_{yz}=33$~\AA, far exceeding the typical range of $10^{-2}$--$1$~\AA\ reported for conventional 2D systems~\cite{PhysRevApplied.13.024053,fuBerryCurvatureDipole2023,jinEnhancedBerryCurvature2021,kimPredictionFerroelectricitydrivenBerry2019d,laiElectricFieldTunable2024a,sharmaLargeBerryCurvature2025}. 

\begin{figure}[!t]
	\centering
	\includegraphics[width=0.45\textwidth]{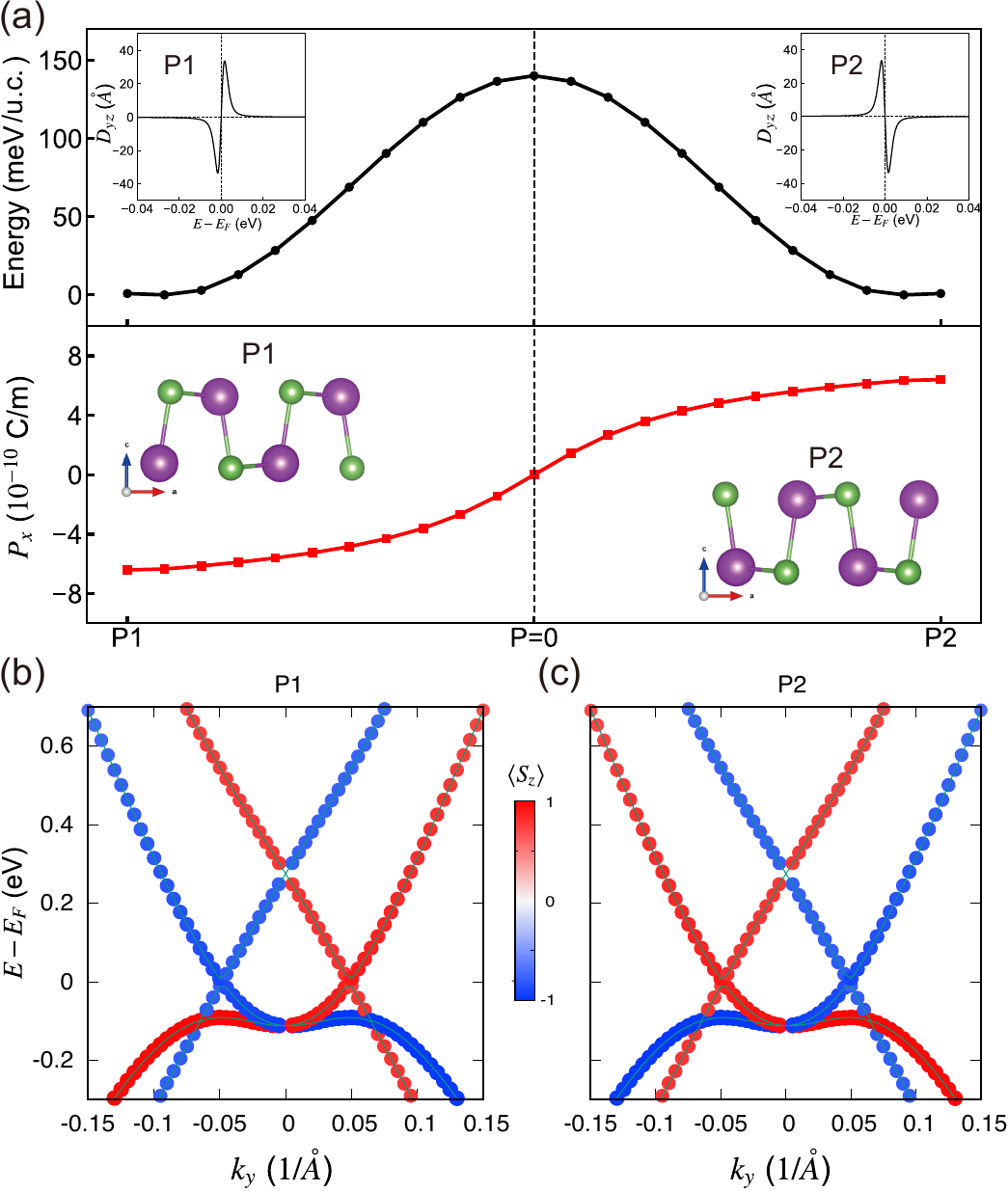}
\caption{
Ferroelectricity and electrical control of the spin texture and BCD in BiAs.
(a) Evolution of the relative energy per unit cell (u.c.) and polarization along the FE switching path. Insets show the BCD $D_{yz}$ for the two opposite FE states (P1 and P2). 
Reversing the FE polarization switches the sign of $D_{yz}$.
(b,c) Ferroelectrically switchable spin texture of spin-Weyl-cone bands. The circle size and color scale represent the magnitude and sign of the spin expectation value $\langle S_z\rangle$.}\label{FE_BCD}
\end{figure}

The large BCD can generate a strong nonlinear charge Hall current and is therefore favorable for terahertz rectification~\cite{zhangTerahertzDetectionBased2021}. To quantify this performance, we evaluate the terahertz current responsivity. Figs.~\ref{BCD_strain}(b) and ~\ref{BCD_strain}(c) show the calculated Drude weight and terahertz current responsivity, respectively. Benefiting from the large intrinsic BCD and small Drude weight near the Weyl nodes, BiAs exhibits a maximum terahertz current responsivity $R_W$ of $3.1~{\rm A/W}$.
This remarkable value of $R_W$ suggests that the intrinsic BiAs monolayer is a promising platform for terahertz rectification. 
As the two $M_y$-related Weyl cones originate from different $S_z$ channels, they also give rise to a huge nonlinear spin Hall conductivity $\chi^z_{yxx}$. The peak value reaches $\chi^z_{yxx}= 1.4~(\hbar/e)~\Omega^{-1} V^{-1}$ [Section~\textcolor{blue}{E} of the SM].

\begin{figure}[!t]
	\centering
	\includegraphics[width=0.45\textwidth]{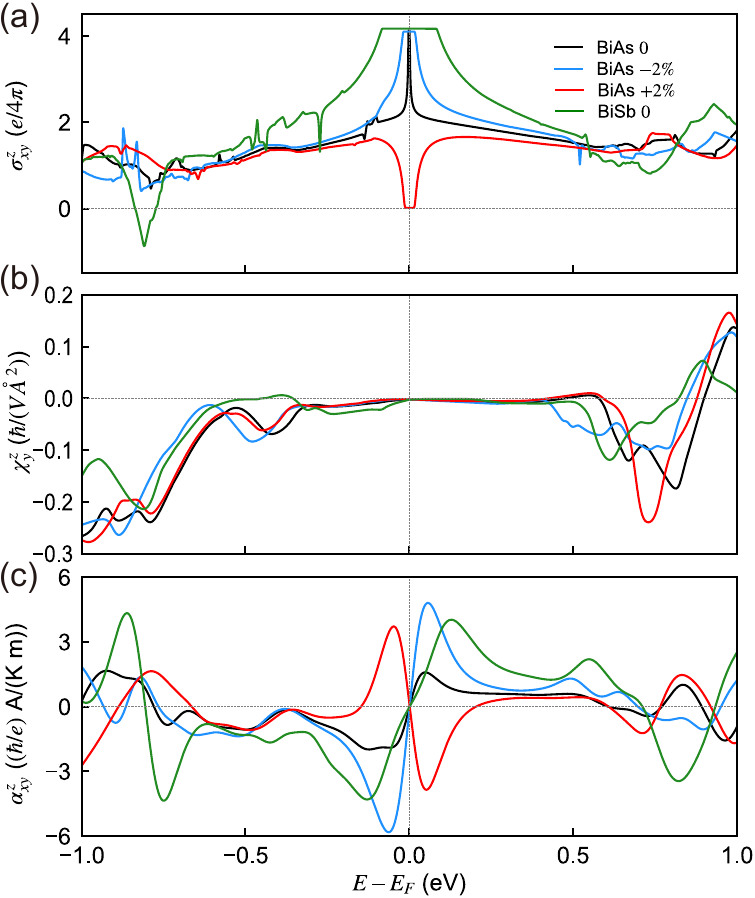}
\caption{
Electrically induced out-of-plane spin accumulation and thermally driven out-of-plane spin transport.
(a) Spin Hall conductivity, (b) Out-of-plane Edelstein response and (c) spin Nernst conductivity as functions of energy $E-E_F$ for BiAs with $\eta=0$, $-2\%$, and $+2\%$, and for BiSb with $\eta=0$. The SHC is calculated using $k_BT=1~{\rm meV}$ ($T=11.6$ K), while the SNC is calculated at $T=300$ K. A spectral broadening of $\Gamma=10$ meV is used in the Edelstein response. The Edelstein response and the SNC are converted to effective bulk values.
}\label{EE_spin_nernst}
\end{figure}

\paragraph*{Tuning by ferroelectric switching.} 
The $\alpha$-Bi-like Group-V puckered-layer family is a ferroelectric.
Fig.~\ref{FE_BCD}(a) shows the energy and polarization evolution of BiAs along the FE switching path. Two inversion-related FE states (P1 and P2) carry opposite $x$-directed polarizations $P_x$.
Since the BCD is odd under inversion, the $D_{yz}$ changes sign upon FE switching, as shown in the inset of Fig.~\ref{FE_BCD}(a). 
Consequently, the rectified terahertz responsivity $R_W$ changes sign. 
In addition, FE reversal can directly switch the $p$-wave valley spin texture, as shown in Figs.~\ref{FE_BCD}(b) and \ref{FE_BCD}(c). These results enable nonvolatile control of spin-polarized transport, such as the Edelstein response.
Beyond electrical control, strain, magnetic fields, and composition can tune the mass of the spin-Weyl cones, BCD, and band topology across the entire $\alpha$-Bi-like Group-V puckered-layer family [Sections~\textcolor{blue}{F},~\textcolor{blue}{G}, and~\textcolor{blue}{H} of the SM].

\paragraph*{Out-of-plane Edelstein effect.}
The Edelstein effect describes electric-field-induced nonequilibrium spin polarization. In coplanar $p$-wave magnets, an in-plane electric field can induce unconventional out-of-plane spin accumulation~\cite{pwave-Chakraborty2025}, while SOC breaks the symmetry protection and allows additional in-plane components.
This may become problematic for devices based on the out-of-plane spin accumulation, when the SOC-induced in-plane susceptibility is comparable to the exchange-driven out-of-plane susceptibility.

By contrast, the fully relativistic crystal symmetry of the nonmagnetic $\alpha$-Bi-like Group-V puckered-layer family allows only the out-of-plane Edelstein component even in the presence of SOC. Within the energy window $E-E_F=\pm0.5$~eV, the symmetry-allowed susceptibility $\chi^z_y$ reaches $0.067~\hbar/(\mathrm{V}\cdot\mathrm{\AA}^2)$ in BiAs and $0.133~\hbar/(\mathrm{V}\cdot\mathrm{\AA}^2)$ in BiSb [Fig.~\ref{EE_spin_nernst}(b)].
FE switching reverses $\chi^z_y$. 
Therefore, the $\alpha$-Bi-like Group-V puckered-layer family can serve as a ferroelectrically switchable out-of-plane spin source for field-free spin-orbit-torque switching in magnetic random-access memory devices~\cite{Kang2025-app}. 
For BiAs, the $\pm2\%$ strains mainly shift the peak positions of $\chi^z_y$ without appreciably changing its magnitude, demonstrating the robustness of the out-of-plane spin accumulation against lattice deformation.

\paragraph*{Spin Nernst effect.}
Beyond electrically induced spin accumulation, the spin-Weyl bands also support thermally driven out-of-plane spin transport. The SHC $\sigma_{xy}^{z}$ as a function of energy $E-E_F$ is shown in Fig.~\ref{EE_spin_nernst}(a).
Since the spin Nernst conductivity (SNC) is a thermal convolution of the SHC, the large SHC and its sharp energy dependence near the critical spin-Weyl transition favor a pronounced spin Nernst response. As shown in Fig.~\ref{EE_spin_nernst}(c), intrinsic BiAs already exhibits a large SNC of about $2~(\hbar/e)$~A/(K$\cdot$m). Under a moderate $-2\%$ strain, the SNC is further enhanced to $5.65~(\hbar/e)$~A/(K$\cdot$m). Intrinsic BiSb also exhibits SNC reaching $4.85~(\hbar/e)$~A/(K$\cdot$m). 
These results demonstrate that spin-Weyl cones provide a route to large and tunable thermally driven spin transport.

\paragraph*{Conclusion and Discussion.}
In conclusion, we establish a general framework for critical spin-Weyl fermions in 2D nonmagnetic systems, and identify the $\alpha$-Bi-like Group-V puckered-layer family of 2D ferroelectrics as a real material platform combining critical spin-Weyl fermions, giant spin splitting, and electrically switchable nonlinear Hall effects and spin transport. 
The tunable tiny mass of the 2D spin-Weyl cones concentrates Berry curvature near the valleys, producing a giant nonlinear charge and spin Hall response. Their unique spin pattern further supports a substantial out-of-plane Edelstein response and a pronounced spin Nernst effect.  
Composition, strain, and magnetic field provide an effective way for tuning the Weyl mass, band topology and associated properties. 
Although the precise band gaps depend on structural details and the exchange-correlation functional [Section~\textcolor{blue}{I} of the SM], strain engineering or chemical composition tuning provides practical routes to critical spin-Weyl nodes within the $\alpha$-Bi family.

Furthermore, ferroelectricity in the BP-like structure offers a key advantage in electrically controlling the $p$-wave spin texture and inversion-odd responses.
Reversing the FE polarization switches the BCD and terahertz rectification signal. Simultaneously, it reverses the valley-resolved spin texture and the out-of-plane Edelstein response.
Through this coupling, the $\alpha$-Bi family offers a nonvolatile and fully electrical route to controlling the nonlinear Hall effect and spin accumulation within a single nonmagnetic FE platform.
In summary, our results establish both a general nonmagnetic
framework and a realistic material platform for exploring critical spin-Weyl physics, nonlinear Hall phenomena, and unconventional spin transport.

\paragraph*{Acknowledgments.}
This work was supported by the National Natural Science Foundation of China (Grant No. 12188101), the National Key R\&D Program of China (Grant No. 2022YFA1403800),  and the Center for Materials Genome. H. W. acknowledges support from the New Cornerstone Science Foundation through the XPLORER PRIZE.


\begin{thebibliography}{42}%
\makeatletter
\providecommand \@ifxundefined [1]{%
 \@ifx{#1\undefined}
}%
\providecommand \@ifnum [1]{%
 \ifnum #1\expandafter \@firstoftwo
 \else \expandafter \@secondoftwo
 \fi
}%
\providecommand \@ifx [1]{%
 \ifx #1\expandafter \@firstoftwo
 \else \expandafter \@secondoftwo
 \fi
}%
\providecommand \natexlab [1]{#1}%
\providecommand \enquote  [1]{``#1''}%
\providecommand \bibnamefont  [1]{#1}%
\providecommand \bibfnamefont [1]{#1}%
\providecommand \citenamefont [1]{#1}%
\providecommand \href@noop [0]{\@secondoftwo}%
\providecommand \href [0]{\begingroup \@sanitize@url \@href}%
\providecommand \@href[1]{\@@startlink{#1}\@@href}%
\providecommand \@@href[1]{\endgroup#1\@@endlink}%
\providecommand \@sanitize@url [0]{\catcode `\\12\catcode `\$12\catcode `\&12\catcode `\#12\catcode `\^12\catcode `\_12\catcode `\%12\relax}%
\providecommand \@@startlink[1]{}%
\providecommand \@@endlink[0]{}%
\providecommand \url  [0]{\begingroup\@sanitize@url \@url }%
\providecommand \@url [1]{\endgroup\@href {#1}{\urlprefix }}%
\providecommand \urlprefix  [0]{URL }%
\providecommand \Eprint [0]{\href }%
\providecommand \doibase [0]{https://doi.org/}%
\providecommand \selectlanguage [0]{\@gobble}%
\providecommand \bibinfo  [0]{\@secondoftwo}%
\providecommand \bibfield  [0]{\@secondoftwo}%
\providecommand \translation [1]{[#1]}%
\providecommand \BibitemOpen [0]{}%
\providecommand \bibitemStop [0]{}%
\providecommand \bibitemNoStop [0]{.\EOS\space}%
\providecommand \EOS [0]{\spacefactor3000\relax}%
\providecommand \BibitemShut  [1]{\csname bibitem#1\endcsname}%
\let\auto@bib@innerbib\@empty
\bibitem [{\citenamefont {Wolf}\ \emph {et~al.}(2001)\citenamefont {Wolf}, \citenamefont {Awschalom}, \citenamefont {Buhrman}, \citenamefont {Daughton}, \citenamefont {von Molnár}, \citenamefont {Roukes}, \citenamefont {Chtchelkanova},\ and\ \citenamefont {Treger}}]{spintronics-doi:10.1126/science.1065389}%
  \BibitemOpen
  \bibfield  {author} {\bibinfo {author} {\bibfnamefont {S.~A.}\ \bibnamefont {Wolf}}, \bibinfo {author} {\bibfnamefont {D.~D.}\ \bibnamefont {Awschalom}}, \bibinfo {author} {\bibfnamefont {R.~A.}\ \bibnamefont {Buhrman}}, \bibinfo {author} {\bibfnamefont {J.~M.}\ \bibnamefont {Daughton}}, \bibinfo {author} {\bibfnamefont {S.}~\bibnamefont {von Molnár}}, \bibinfo {author} {\bibfnamefont {M.~L.}\ \bibnamefont {Roukes}}, \bibinfo {author} {\bibfnamefont {A.~Y.}\ \bibnamefont {Chtchelkanova}},\ and\ \bibinfo {author} {\bibfnamefont {D.~M.}\ \bibnamefont {Treger}},\ }\bibfield  {title} {\bibinfo {title} {Spintronics: A spin-based electronics vision for the future},\ }\href {https://doi.org/10.1126/science.1065389} {\bibfield  {journal} {\bibinfo  {journal} {Science}\ }\textbf {\bibinfo {volume} {294}},\ \bibinfo {pages} {1488} (\bibinfo {year} {2001})}\BibitemShut {NoStop}%
\bibitem [{\citenamefont {\ifmmode \check{Z}\else \v{Z}\fi{}uti\ifmmode~\acute{c}\else \'{c}\fi{}}\ \emph {et~al.}(2004)\citenamefont {\ifmmode \check{Z}\else \v{Z}\fi{}uti\ifmmode~\acute{c}\else \'{c}\fi{}}, \citenamefont {Fabian},\ and\ \citenamefont {Das~Sarma}}]{spintronics-RevModPhys.76.323}%
  \BibitemOpen
  \bibfield  {author} {\bibinfo {author} {\bibfnamefont {I.}~\bibnamefont {\ifmmode \check{Z}\else \v{Z}\fi{}uti\ifmmode~\acute{c}\else \'{c}\fi{}}}, \bibinfo {author} {\bibfnamefont {J.}~\bibnamefont {Fabian}},\ and\ \bibinfo {author} {\bibfnamefont {S.}~\bibnamefont {Das~Sarma}},\ }\bibfield  {title} {\bibinfo {title} {Spintronics: Fundamentals and applications},\ }\href {https://doi.org/10.1103/RevModPhys.76.323} {\bibfield  {journal} {\bibinfo  {journal} {Rev. Mod. Phys.}\ }\textbf {\bibinfo {volume} {76}},\ \bibinfo {pages} {323} (\bibinfo {year} {2004})}\BibitemShut {NoStop}%
\bibitem [{\citenamefont {Awschalom}\ and\ \citenamefont {Flatt{\'e}}(2007)}]{spintronics-Awschalom2007}%
  \BibitemOpen
  \bibfield  {author} {\bibinfo {author} {\bibfnamefont {D.~D.}\ \bibnamefont {Awschalom}}\ and\ \bibinfo {author} {\bibfnamefont {M.~E.}\ \bibnamefont {Flatt{\'e}}},\ }\bibfield  {title} {\bibinfo {title} {Challenges for semiconductor spintronics},\ }\href {https://doi.org/10.1038/nphys551} {\bibfield  {journal} {\bibinfo  {journal} {Nature Physics}\ }\textbf {\bibinfo {volume} {3}},\ \bibinfo {pages} {153} (\bibinfo {year} {2007})}\BibitemShut {NoStop}%
\bibitem [{\citenamefont {Slater}(1936)}]{FMspin-PhysRev.49.537}%
  \BibitemOpen
  \bibfield  {author} {\bibinfo {author} {\bibfnamefont {J.~C.}\ \bibnamefont {Slater}},\ }\bibfield  {title} {\bibinfo {title} {The ferromagnetism of nickel},\ }\href {https://doi.org/10.1103/PhysRev.49.537} {\bibfield  {journal} {\bibinfo  {journal} {Phys. Rev.}\ }\textbf {\bibinfo {volume} {49}},\ \bibinfo {pages} {537} (\bibinfo {year} {1936})}\BibitemShut {NoStop}%
\bibitem [{\citenamefont {Stoner}(1938)}]{FMspin-Stoner1938}%
  \BibitemOpen
  \bibfield  {author} {\bibinfo {author} {\bibfnamefont {E.~C.}\ \bibnamefont {Stoner}},\ }\bibfield  {title} {\bibinfo {title} {Collective electron ferromagnetism},\ }\href {https://doi.org/10.1098/rspa.1938.0066} {\bibfield  {journal} {\bibinfo  {journal} {Proceedings of the Royal Society of London. A. Mathematical and Physical Sciences}\ }\textbf {\bibinfo {volume} {165}},\ \bibinfo {pages} {372} (\bibinfo {year} {1938})}\BibitemShut {NoStop}%
\bibitem [{\citenamefont {Liu}\ \emph {et~al.}(2022)\citenamefont {Liu}, \citenamefont {Zhang}, \citenamefont {Nie}, \citenamefont {Liu}, \citenamefont {Sun}, \citenamefont {Wang}, \citenamefont {Ding}, \citenamefont {Jiang}, \citenamefont {Sun}, \citenamefont {Xue}, \citenamefont {Huang}, \citenamefont {Su}, \citenamefont {Yang}, \citenamefont {Jiang}, \citenamefont {Lu}, \citenamefont {Yuan}, \citenamefont {Cho}, \citenamefont {Liu}, \citenamefont {Liu}, \citenamefont {Ye}, \citenamefont {Zhang}, \citenamefont {Weng}, \citenamefont {Liu}, \citenamefont {Guo}, \citenamefont {Wang},\ and\ \citenamefont {Shen}}]{liuSpontaneousFerromagnetismInduced2022}%
  \BibitemOpen
  \bibfield  {author} {\bibinfo {author} {\bibfnamefont {W.~L.}\ \bibnamefont {Liu}}, \bibinfo {author} {\bibfnamefont {X.}~\bibnamefont {Zhang}}, \bibinfo {author} {\bibfnamefont {S.~M.}\ \bibnamefont {Nie}}, \bibinfo {author} {\bibfnamefont {Z.~T.}\ \bibnamefont {Liu}}, \bibinfo {author} {\bibfnamefont {X.~Y.}\ \bibnamefont {Sun}}, \bibinfo {author} {\bibfnamefont {H.~Y.}\ \bibnamefont {Wang}}, \bibinfo {author} {\bibfnamefont {J.~Y.}\ \bibnamefont {Ding}}, \bibinfo {author} {\bibfnamefont {Q.}~\bibnamefont {Jiang}}, \bibinfo {author} {\bibfnamefont {L.}~\bibnamefont {Sun}}, \bibinfo {author} {\bibfnamefont {F.~H.}\ \bibnamefont {Xue}}, \bibinfo {author} {\bibfnamefont {Z.}~\bibnamefont {Huang}}, \bibinfo {author} {\bibfnamefont {H.}~\bibnamefont {Su}}, \bibinfo {author} {\bibfnamefont {Y.~C.}\ \bibnamefont {Yang}}, \bibinfo {author} {\bibfnamefont {Z.~C.}\ \bibnamefont {Jiang}}, \bibinfo {author} {\bibfnamefont {X.~L.}\ \bibnamefont {Lu}}, \bibinfo {author} {\bibfnamefont {J.}~\bibnamefont {Yuan}},
  \bibinfo {author} {\bibfnamefont {S.}~\bibnamefont {Cho}}, \bibinfo {author} {\bibfnamefont {J.~S.}\ \bibnamefont {Liu}}, \bibinfo {author} {\bibfnamefont {Z.~H.}\ \bibnamefont {Liu}}, \bibinfo {author} {\bibfnamefont {M.}~\bibnamefont {Ye}}, \bibinfo {author} {\bibfnamefont {S.~L.}\ \bibnamefont {Zhang}}, \bibinfo {author} {\bibfnamefont {H.~M.}\ \bibnamefont {Weng}}, \bibinfo {author} {\bibfnamefont {Z.}~\bibnamefont {Liu}}, \bibinfo {author} {\bibfnamefont {Y.~F.}\ \bibnamefont {Guo}}, \bibinfo {author} {\bibfnamefont {Z.~J.}\ \bibnamefont {Wang}},\ and\ \bibinfo {author} {\bibfnamefont {D.~W.}\ \bibnamefont {Shen}},\ }\bibfield  {title} {\bibinfo {title} {Spontaneous {{Ferromagnetism Induced Topological Transition}} in {{EuB}} 6},\ }\href {https://doi.org/10.1103/PhysRevLett.129.166402} {\bibfield  {journal} {\bibinfo  {journal} {Phys. Rev. Lett.}\ }\textbf {\bibinfo {volume} {129}},\ \bibinfo {pages} {166402} (\bibinfo {year} {2022})}\BibitemShut {NoStop}%
\bibitem [{\citenamefont {Yao}\ \emph {et~al.}(2025)\citenamefont {Yao}, \citenamefont {Zhang}, \citenamefont {Zhang}, \citenamefont {Sheng}, \citenamefont {Shi}, \citenamefont {Fang}, \citenamefont {Weng},\ and\ \citenamefont {Wang}}]{yaoAxionInsulatorWeyl2025}%
  \BibitemOpen
  \bibfield  {author} {\bibinfo {author} {\bibfnamefont {J.}~\bibnamefont {Yao}}, \bibinfo {author} {\bibfnamefont {R.}~\bibnamefont {Zhang}}, \bibinfo {author} {\bibfnamefont {S.}~\bibnamefont {Zhang}}, \bibinfo {author} {\bibfnamefont {H.}~\bibnamefont {Sheng}}, \bibinfo {author} {\bibfnamefont {Y.}~\bibnamefont {Shi}}, \bibinfo {author} {\bibfnamefont {Z.}~\bibnamefont {Fang}}, \bibinfo {author} {\bibfnamefont {H.}~\bibnamefont {Weng}},\ and\ \bibinfo {author} {\bibfnamefont {Z.}~\bibnamefont {Wang}},\ }\bibfield  {title} {\bibinfo {title} {Axion insulator, {{Weyl}} points, quantum anomalous {{Hall}} effect, and magnetic topological phase transition in {{Eu}} 3 {{In}} 2 {{As}} 4},\ }\href {https://doi.org/10.1103/PhysRevB.111.L041117} {\bibfield  {journal} {\bibinfo  {journal} {Phys. Rev. B}\ }\textbf {\bibinfo {volume} {111}},\ \bibinfo {pages} {L041117} (\bibinfo {year} {2025})}\BibitemShut {NoStop}%
\bibitem [{\citenamefont {Dresselhaus}(1955)}]{socspin-PhysRev.100.580}%
  \BibitemOpen
  \bibfield  {author} {\bibinfo {author} {\bibfnamefont {G.}~\bibnamefont {Dresselhaus}},\ }\bibfield  {title} {\bibinfo {title} {Spin-orbit coupling effects in zinc blende structures},\ }\href {https://doi.org/10.1103/PhysRev.100.580} {\bibfield  {journal} {\bibinfo  {journal} {Phys. Rev.}\ }\textbf {\bibinfo {volume} {100}},\ \bibinfo {pages} {580} (\bibinfo {year} {1955})}\BibitemShut {NoStop}%
\bibitem [{\citenamefont {Bihlmayer}\ \emph {et~al.}(2022)\citenamefont {Bihlmayer}, \citenamefont {No{\"e}l}, \citenamefont {Vyalikh}, \citenamefont {Chulkov},\ and\ \citenamefont {Manchon}}]{socspin-Bihlmayer2022}%
  \BibitemOpen
  \bibfield  {author} {\bibinfo {author} {\bibfnamefont {G.}~\bibnamefont {Bihlmayer}}, \bibinfo {author} {\bibfnamefont {P.}~\bibnamefont {No{\"e}l}}, \bibinfo {author} {\bibfnamefont {D.~V.}\ \bibnamefont {Vyalikh}}, \bibinfo {author} {\bibfnamefont {E.~V.}\ \bibnamefont {Chulkov}},\ and\ \bibinfo {author} {\bibfnamefont {A.}~\bibnamefont {Manchon}},\ }\bibfield  {title} {\bibinfo {title} {Rashba-like physics in condensed matter},\ }\href {https://doi.org/10.1038/s42254-022-00490-y} {\bibfield  {journal} {\bibinfo  {journal} {Nature Reviews Physics}\ }\textbf {\bibinfo {volume} {4}},\ \bibinfo {pages} {642} (\bibinfo {year} {2022})}\BibitemShut {NoStop}%
\bibitem [{\citenamefont {Shao}\ \emph {et~al.}(2023)\citenamefont {Shao}, \citenamefont {Deng}, \citenamefont {Sheng}, \citenamefont {Zhang}, \citenamefont {Weng}, \citenamefont {Fang}, \citenamefont {Chen}, \citenamefont {Sun},\ and\ \citenamefont {Wang}}]{shaoLargeSpinHall2023}%
  \BibitemOpen
  \bibfield  {author} {\bibinfo {author} {\bibfnamefont {D.}~\bibnamefont {Shao}}, \bibinfo {author} {\bibfnamefont {J.}~\bibnamefont {Deng}}, \bibinfo {author} {\bibfnamefont {H.}~\bibnamefont {Sheng}}, \bibinfo {author} {\bibfnamefont {R.}~\bibnamefont {Zhang}}, \bibinfo {author} {\bibfnamefont {H.}~\bibnamefont {Weng}}, \bibinfo {author} {\bibfnamefont {Z.}~\bibnamefont {Fang}}, \bibinfo {author} {\bibfnamefont {X.-Q.}\ \bibnamefont {Chen}}, \bibinfo {author} {\bibfnamefont {Y.}~\bibnamefont {Sun}},\ and\ \bibinfo {author} {\bibfnamefont {Z.}~\bibnamefont {Wang}},\ }\bibfield  {title} {\bibinfo {title} {Large {{Spin Hall Conductivity}} and {{Excellent Hydrogen Evolution Reaction Activity}} in {{Unconventional PtTe}}{\textsubscript{1.75}} {{Monolayer}}},\ }\href {https://doi.org/10.34133/research.0042} {\bibfield  {journal} {\bibinfo  {journal} {Research}\ }\textbf {\bibinfo {volume} {6}},\ \bibinfo {pages} {0042} (\bibinfo {year} {2023})}\BibitemShut {NoStop}%
\bibitem [{\citenamefont {Sch{\"a}fer}(2021)}]{Schaefer2021Domains}%
  \BibitemOpen
  \bibfield  {author} {\bibinfo {author} {\bibfnamefont {R.}~\bibnamefont {Sch{\"a}fer}},\ }\bibfield  {title} {\bibinfo {title} {Magnetic domains},\ }in\ \href {https://doi.org/10.1007/978-3-030-63210-6\_8} {\emph {\bibinfo {booktitle} {Handbook of Magnetism and Magnetic Materials}}},\ \bibinfo {editor} {edited by\ \bibinfo {editor} {\bibfnamefont {J.~M.~D.}\ \bibnamefont {Coey}}\ and\ \bibinfo {editor} {\bibfnamefont {S.~S.~P.}\ \bibnamefont {Parkin}}}\ (\bibinfo  {publisher} {Springer},\ \bibinfo {address} {Cham},\ \bibinfo {year} {2021})\ pp.\ \bibinfo {pages} {391--434}\BibitemShut {NoStop}%
\bibitem [{New(2022)}]{NewHorizonsSpintronics2022}%
  \BibitemOpen
  \bibfield  {title} {\bibinfo {title} {New horizons in spintronics},\ }\href {https://doi.org/10.1038/s41563-021-01184-z} {\bibfield  {journal} {\bibinfo  {journal} {Nature Materials}\ }\textbf {\bibinfo {volume} {21}},\ \bibinfo {pages} {1} (\bibinfo {year} {2022})}\BibitemShut {NoStop}%
\bibitem [{\citenamefont {Matsukura}\ \emph {et~al.}(2015)\citenamefont {Matsukura}, \citenamefont {Tokura},\ and\ \citenamefont {Ohno}}]{Matsukura2015NNano}%
  \BibitemOpen
  \bibfield  {author} {\bibinfo {author} {\bibfnamefont {F.}~\bibnamefont {Matsukura}}, \bibinfo {author} {\bibfnamefont {Y.}~\bibnamefont {Tokura}},\ and\ \bibinfo {author} {\bibfnamefont {H.}~\bibnamefont {Ohno}},\ }\bibfield  {title} {\bibinfo {title} {Control of magnetism by electric fields},\ }\href {https://doi.org/10.1038/nnano.2015.22} {\bibfield  {journal} {\bibinfo  {journal} {Nature Nanotechnology}\ }\textbf {\bibinfo {volume} {10}},\ \bibinfo {pages} {209} (\bibinfo {year} {2015})}\BibitemShut {NoStop}%
\bibitem [{\citenamefont {Liu}\ \emph {et~al.}(2024)\citenamefont {Liu}, \citenamefont {Cao}, \citenamefont {Zhang}, \citenamefont {Liang}, \citenamefont {Wang},\ and\ \citenamefont {Yang}}]{Liu2024PRB}%
  \BibitemOpen
  \bibfield  {author} {\bibinfo {author} {\bibfnamefont {H.}~\bibnamefont {Liu}}, \bibinfo {author} {\bibfnamefont {J.}~\bibnamefont {Cao}}, \bibinfo {author} {\bibfnamefont {Z.}~\bibnamefont {Zhang}}, \bibinfo {author} {\bibfnamefont {J.}~\bibnamefont {Liang}}, \bibinfo {author} {\bibfnamefont {L.}~\bibnamefont {Wang}},\ and\ \bibinfo {author} {\bibfnamefont {S.~A.}\ \bibnamefont {Yang}},\ }\bibfield  {title} {\bibinfo {title} {Ideal spin-polarized weyl half-semimetal with a single pair of weyl points in half-heusler compounds {XCrTe} ({X} = {K}, {Rb})},\ }\href {https://doi.org/10.1103/PhysRevB.109.174426} {\bibfield  {journal} {\bibinfo  {journal} {Phys. Rev. B}\ }\textbf {\bibinfo {volume} {109}},\ \bibinfo {pages} {174426} (\bibinfo {year} {2024})}\BibitemShut {NoStop}%
\bibitem [{\citenamefont {Li}\ \emph {et~al.}(2025)\citenamefont {Li}, \citenamefont {Chen}, \citenamefont {Zhang},\ and\ \citenamefont {Fu}}]{Li2N-2025-vqdz-12bz}%
  \BibitemOpen
  \bibfield  {author} {\bibinfo {author} {\bibfnamefont {Q.}~\bibnamefont {Li}}, \bibinfo {author} {\bibfnamefont {L.}~\bibnamefont {Chen}}, \bibinfo {author} {\bibfnamefont {R.-W.}\ \bibnamefont {Zhang}},\ and\ \bibinfo {author} {\bibfnamefont {B.}~\bibnamefont {Fu}},\ }\bibfield  {title} {\bibinfo {title} {Type-iii weyl semi-half-metal in an ultralight monolayer ${\mathrm{li}}_{2}\mathrm{N}$},\ }\href {https://doi.org/10.1103/vqdz-12bz} {\bibfield  {journal} {\bibinfo  {journal} {Phys. Rev. B}\ }\textbf {\bibinfo {volume} {112}},\ \bibinfo {pages} {195126} (\bibinfo {year} {2025})}\BibitemShut {NoStop}%
\bibitem [{\citenamefont {Bychkov}\ and\ \citenamefont {Rashba}(1984)}]{Bychkov1984JETPL}%
  \BibitemOpen
  \bibfield  {author} {\bibinfo {author} {\bibfnamefont {Y.~A.}\ \bibnamefont {Bychkov}}\ and\ \bibinfo {author} {\bibfnamefont {E.~I.}\ \bibnamefont {Rashba}},\ }\bibfield  {title} {\bibinfo {title} {Properties of a two-dimensional electron gas with lifted spectral degeneracy},\ }\href@noop {} {\bibfield  {journal} {\bibinfo  {journal} {JETP Letters}\ }\textbf {\bibinfo {volume} {39}},\ \bibinfo {pages} {78} (\bibinfo {year} {1984})}\BibitemShut {NoStop}%
\bibitem [{\citenamefont {Hsieh}\ \emph {et~al.}(2009)\citenamefont {Hsieh}, \citenamefont {Xia}, \citenamefont {Wray}, \citenamefont {Qian}, \citenamefont {Pal}, \citenamefont {Dil}, \citenamefont {Osterwalder}, \citenamefont {Meier}, \citenamefont {Bihlmayer}, \citenamefont {Kane}, \citenamefont {Hor}, \citenamefont {Cava},\ and\ \citenamefont {Hasan}}]{Hsieh2009Science}%
  \BibitemOpen
  \bibfield  {author} {\bibinfo {author} {\bibfnamefont {D.}~\bibnamefont {Hsieh}}, \bibinfo {author} {\bibfnamefont {Y.}~\bibnamefont {Xia}}, \bibinfo {author} {\bibfnamefont {L.}~\bibnamefont {Wray}}, \bibinfo {author} {\bibfnamefont {D.}~\bibnamefont {Qian}}, \bibinfo {author} {\bibfnamefont {A.}~\bibnamefont {Pal}}, \bibinfo {author} {\bibfnamefont {J.~H.}\ \bibnamefont {Dil}}, \bibinfo {author} {\bibfnamefont {J.}~\bibnamefont {Osterwalder}}, \bibinfo {author} {\bibfnamefont {F.}~\bibnamefont {Meier}}, \bibinfo {author} {\bibfnamefont {G.}~\bibnamefont {Bihlmayer}}, \bibinfo {author} {\bibfnamefont {C.~L.}\ \bibnamefont {Kane}}, \bibinfo {author} {\bibfnamefont {Y.~S.}\ \bibnamefont {Hor}}, \bibinfo {author} {\bibfnamefont {R.~J.}\ \bibnamefont {Cava}},\ and\ \bibinfo {author} {\bibfnamefont {M.~Z.}\ \bibnamefont {Hasan}},\ }\bibfield  {title} {\bibinfo {title} {Observation of unconventional quantum spin textures in topological insulators},\ }\href {https://doi.org/10.1126/science.1167733} {\bibfield
  {journal} {\bibinfo  {journal} {Science}\ }\textbf {\bibinfo {volume} {323}},\ \bibinfo {pages} {919} (\bibinfo {year} {2009})}\BibitemShut {NoStop}%
\bibitem [{\citenamefont {Xiao}\ \emph {et~al.}(2007)\citenamefont {Xiao}, \citenamefont {Yao},\ and\ \citenamefont {Niu}}]{Xiao2007PRL}%
  \BibitemOpen
  \bibfield  {author} {\bibinfo {author} {\bibfnamefont {D.}~\bibnamefont {Xiao}}, \bibinfo {author} {\bibfnamefont {W.}~\bibnamefont {Yao}},\ and\ \bibinfo {author} {\bibfnamefont {Q.}~\bibnamefont {Niu}},\ }\bibfield  {title} {\bibinfo {title} {Valley-contrasting physics in graphene: Magnetic moment and topological transport},\ }\href {https://doi.org/10.1103/PhysRevLett.99.236809} {\bibfield  {journal} {\bibinfo  {journal} {Phys. Rev. Lett.}\ }\textbf {\bibinfo {volume} {99}},\ \bibinfo {pages} {236809} (\bibinfo {year} {2007})}\BibitemShut {NoStop}%
\bibitem [{\citenamefont {\ifmmode~\check{S}\else \v{S}\fi{}mejkal}\ \emph {et~al.}(2022{\natexlab{a}})\citenamefont {\ifmmode~\check{S}\else \v{S}\fi{}mejkal}, \citenamefont {Hellenes}, \citenamefont {Gonz\'alez-Hern\'andez}, \citenamefont {Sinova},\ and\ \citenamefont {Jungwirth}}]{unConvenmag-PhysRevX.12.011028}%
  \BibitemOpen
  \bibfield  {author} {\bibinfo {author} {\bibfnamefont {L.}~\bibnamefont {\ifmmode~\check{S}\else \v{S}\fi{}mejkal}}, \bibinfo {author} {\bibfnamefont {A.~B.}\ \bibnamefont {Hellenes}}, \bibinfo {author} {\bibfnamefont {R.}~\bibnamefont {Gonz\'alez-Hern\'andez}}, \bibinfo {author} {\bibfnamefont {J.}~\bibnamefont {Sinova}},\ and\ \bibinfo {author} {\bibfnamefont {T.}~\bibnamefont {Jungwirth}},\ }\bibfield  {title} {\bibinfo {title} {Giant and tunneling magnetoresistance in unconventional collinear antiferromagnets with nonrelativistic spin-momentum coupling},\ }\href {https://doi.org/10.1103/PhysRevX.12.011028} {\bibfield  {journal} {\bibinfo  {journal} {Phys. Rev. X}\ }\textbf {\bibinfo {volume} {12}},\ \bibinfo {pages} {011028} (\bibinfo {year} {2022}{\natexlab{a}})}\BibitemShut {NoStop}%
\bibitem [{\citenamefont {\ifmmode~\check{S}\else \v{S}\fi{}mejkal}\ \emph {et~al.}(2022{\natexlab{b}})\citenamefont {\ifmmode~\check{S}\else \v{S}\fi{}mejkal}, \citenamefont {Sinova},\ and\ \citenamefont {Jungwirth}}]{unConvenmag-PhysRevX.12.031042}%
  \BibitemOpen
  \bibfield  {author} {\bibinfo {author} {\bibfnamefont {L.}~\bibnamefont {\ifmmode~\check{S}\else \v{S}\fi{}mejkal}}, \bibinfo {author} {\bibfnamefont {J.}~\bibnamefont {Sinova}},\ and\ \bibinfo {author} {\bibfnamefont {T.}~\bibnamefont {Jungwirth}},\ }\bibfield  {title} {\bibinfo {title} {Beyond conventional ferromagnetism and antiferromagnetism: A phase with nonrelativistic spin and crystal rotation symmetry},\ }\href {https://doi.org/10.1103/PhysRevX.12.031042} {\bibfield  {journal} {\bibinfo  {journal} {Phys. Rev. X}\ }\textbf {\bibinfo {volume} {12}},\ \bibinfo {pages} {031042} (\bibinfo {year} {2022}{\natexlab{b}})}\BibitemShut {NoStop}%
\bibitem [{\citenamefont {\ifmmode~\check{S}\else \v{S}\fi{}mejkal}\ \emph {et~al.}(2022{\natexlab{c}})\citenamefont {\ifmmode~\check{S}\else \v{S}\fi{}mejkal}, \citenamefont {Sinova},\ and\ \citenamefont {Jungwirth}}]{unConvenmag-PhysRevX.12.040501}%
  \BibitemOpen
  \bibfield  {author} {\bibinfo {author} {\bibfnamefont {L.}~\bibnamefont {\ifmmode~\check{S}\else \v{S}\fi{}mejkal}}, \bibinfo {author} {\bibfnamefont {J.}~\bibnamefont {Sinova}},\ and\ \bibinfo {author} {\bibfnamefont {T.}~\bibnamefont {Jungwirth}},\ }\bibfield  {title} {\bibinfo {title} {Emerging research landscape of altermagnetism},\ }\href {https://doi.org/10.1103/PhysRevX.12.040501} {\bibfield  {journal} {\bibinfo  {journal} {Phys. Rev. X}\ }\textbf {\bibinfo {volume} {12}},\ \bibinfo {pages} {040501} (\bibinfo {year} {2022}{\natexlab{c}})}\BibitemShut {NoStop}%
\bibitem [{\citenamefont {Song}\ \emph {et~al.}(2025)\citenamefont {Song}, \citenamefont {Qi}, \citenamefont {Fang}, \citenamefont {Fang},\ and\ \citenamefont {Weng}}]{ssg-szy-song2025unifiedsymmetryclassificationmagnetic}%
  \BibitemOpen
  \bibfield  {author} {\bibinfo {author} {\bibfnamefont {Z.}~\bibnamefont {Song}}, \bibinfo {author} {\bibfnamefont {Z.}~\bibnamefont {Qi}}, \bibinfo {author} {\bibfnamefont {C.}~\bibnamefont {Fang}}, \bibinfo {author} {\bibfnamefont {Z.}~\bibnamefont {Fang}},\ and\ \bibinfo {author} {\bibfnamefont {H.}~\bibnamefont {Weng}},\ }\href {https://arxiv.org/abs/2512.08901} {\bibinfo {title} {A unified symmetry classification of magnetic orders via spin space groups: Prediction of coplanar even-wave phases}} (\bibinfo {year} {2025}),\ \Eprint {https://arxiv.org/abs/2512.08901} {arXiv:2512.08901 [cond-mat.mtrl-sci]} \BibitemShut {NoStop}%
\bibitem [{\citenamefont {Zhang}\ \emph {et~al.}(2026)\citenamefont {Zhang}, \citenamefont {Song}, \citenamefont {Fang}, \citenamefont {Weng},\ and\ \citenamefont {Wang}}]{irssg}%
  \BibitemOpen
  \bibfield  {author} {\bibinfo {author} {\bibfnamefont {S.}~\bibnamefont {Zhang}}, \bibinfo {author} {\bibfnamefont {Z.}~\bibnamefont {Song}}, \bibinfo {author} {\bibfnamefont {Z.}~\bibnamefont {Fang}}, \bibinfo {author} {\bibfnamefont {H.}~\bibnamefont {Weng}},\ and\ \bibinfo {author} {\bibfnamefont {Z.}~\bibnamefont {Wang}},\ }\bibfield  {title} {\bibinfo {title} {Irssg: An open-source software package for spin space groups},\ }\href {https://doi.org/https://doi.org/10.1016/j.cpc.2026.110190} {\bibfield  {journal} {\bibinfo  {journal} {Computer Physics Communications}\ }\textbf {\bibinfo {volume} {326}},\ \bibinfo {pages} {110190} (\bibinfo {year} {2026})}\BibitemShut {NoStop}%
\bibitem [{\citenamefont {Hellenes}\ \emph {et~al.}(2024)\citenamefont {Hellenes}, \citenamefont {Jungwirth}, \citenamefont {Jaeschke-Ubiergo}, \citenamefont {Chakraborty}, \citenamefont {Sinova},\ and\ \citenamefont {Šmejkal}}]{pwave-hellenes2024pwavemagnets}%
  \BibitemOpen
  \bibfield  {author} {\bibinfo {author} {\bibfnamefont {A.~B.}\ \bibnamefont {Hellenes}}, \bibinfo {author} {\bibfnamefont {T.}~\bibnamefont {Jungwirth}}, \bibinfo {author} {\bibfnamefont {R.}~\bibnamefont {Jaeschke-Ubiergo}}, \bibinfo {author} {\bibfnamefont {A.}~\bibnamefont {Chakraborty}}, \bibinfo {author} {\bibfnamefont {J.}~\bibnamefont {Sinova}},\ and\ \bibinfo {author} {\bibfnamefont {L.}~\bibnamefont {Šmejkal}},\ }\href {https://arxiv.org/abs/2309.01607} {\bibinfo {title} {P-wave magnets}} (\bibinfo {year} {2024}),\ \Eprint {https://arxiv.org/abs/2309.01607} {arXiv:2309.01607 [cond-mat.mes-hall]} \BibitemShut {NoStop}%
\bibitem [{\citenamefont {Yamada}\ \emph {et~al.}(2025)\citenamefont {Yamada}, \citenamefont {Birch}, \citenamefont {Baral}, \citenamefont {Okumura}, \citenamefont {Nakano}, \citenamefont {Gao}, \citenamefont {Ezawa}, \citenamefont {Nomoto}, \citenamefont {Masell}, \citenamefont {Ishihara}, \citenamefont {Kolincio}, \citenamefont {Belopolski}, \citenamefont {Sagayama}, \citenamefont {Nakao}, \citenamefont {Ohishi}, \citenamefont {Ohhara}, \citenamefont {Kiyanagi}, \citenamefont {Nakajima}, \citenamefont {Tokura}, \citenamefont {Arima}, \citenamefont {Motome}, \citenamefont {Hirschmann},\ and\ \citenamefont {Hirschberger}}]{pwave-Yamada2025}%
  \BibitemOpen
  \bibfield  {author} {\bibinfo {author} {\bibfnamefont {R.}~\bibnamefont {Yamada}}, \bibinfo {author} {\bibfnamefont {M.~T.}\ \bibnamefont {Birch}}, \bibinfo {author} {\bibfnamefont {P.~R.}\ \bibnamefont {Baral}}, \bibinfo {author} {\bibfnamefont {S.}~\bibnamefont {Okumura}}, \bibinfo {author} {\bibfnamefont {R.}~\bibnamefont {Nakano}}, \bibinfo {author} {\bibfnamefont {S.}~\bibnamefont {Gao}}, \bibinfo {author} {\bibfnamefont {M.}~\bibnamefont {Ezawa}}, \bibinfo {author} {\bibfnamefont {T.}~\bibnamefont {Nomoto}}, \bibinfo {author} {\bibfnamefont {J.}~\bibnamefont {Masell}}, \bibinfo {author} {\bibfnamefont {Y.}~\bibnamefont {Ishihara}}, \bibinfo {author} {\bibfnamefont {K.~K.}\ \bibnamefont {Kolincio}}, \bibinfo {author} {\bibfnamefont {I.}~\bibnamefont {Belopolski}}, \bibinfo {author} {\bibfnamefont {H.}~\bibnamefont {Sagayama}}, \bibinfo {author} {\bibfnamefont {H.}~\bibnamefont {Nakao}}, \bibinfo {author} {\bibfnamefont {K.}~\bibnamefont {Ohishi}}, \bibinfo {author} {\bibfnamefont {T.}~\bibnamefont
  {Ohhara}}, \bibinfo {author} {\bibfnamefont {R.}~\bibnamefont {Kiyanagi}}, \bibinfo {author} {\bibfnamefont {T.}~\bibnamefont {Nakajima}}, \bibinfo {author} {\bibfnamefont {Y.}~\bibnamefont {Tokura}}, \bibinfo {author} {\bibfnamefont {T.-h.}\ \bibnamefont {Arima}}, \bibinfo {author} {\bibfnamefont {Y.}~\bibnamefont {Motome}}, \bibinfo {author} {\bibfnamefont {M.~M.}\ \bibnamefont {Hirschmann}},\ and\ \bibinfo {author} {\bibfnamefont {M.}~\bibnamefont {Hirschberger}},\ }\bibfield  {title} {\bibinfo {title} {A metallic {p}-wave magnet with commensurate spin helix},\ }\href {https://doi.org/10.1038/s41586-025-09633-4} {\bibfield  {journal} {\bibinfo  {journal} {Nature}\ }\textbf {\bibinfo {volume} {646}},\ \bibinfo {pages} {837} (\bibinfo {year} {2025})}\BibitemShut {NoStop}%
\bibitem [{\citenamefont {Chakraborty}\ \emph {et~al.}(2025)\citenamefont {Chakraborty}, \citenamefont {Birk~Hellenes}, \citenamefont {Jaeschke-Ubiergo}, \citenamefont {Jungwirth}, \citenamefont {{\v S}mejkal},\ and\ \citenamefont {Sinova}}]{pwave-Chakraborty2025}%
  \BibitemOpen
  \bibfield  {author} {\bibinfo {author} {\bibfnamefont {A.}~\bibnamefont {Chakraborty}}, \bibinfo {author} {\bibfnamefont {A.}~\bibnamefont {Birk~Hellenes}}, \bibinfo {author} {\bibfnamefont {R.}~\bibnamefont {Jaeschke-Ubiergo}}, \bibinfo {author} {\bibfnamefont {T.}~\bibnamefont {Jungwirth}}, \bibinfo {author} {\bibfnamefont {L.}~\bibnamefont {{\v S}mejkal}},\ and\ \bibinfo {author} {\bibfnamefont {J.}~\bibnamefont {Sinova}},\ }\bibfield  {title} {\bibinfo {title} {Highly efficient non-relativistic {Edelstein} effect in nodal {p}-wave magnets},\ }\href {https://doi.org/10.1038/s41467-025-62516-0} {\bibfield  {journal} {\bibinfo  {journal} {Nature Communications}\ }\textbf {\bibinfo {volume} {16}},\ \bibinfo {pages} {7270} (\bibinfo {year} {2025})}\BibitemShut {NoStop}%
\bibitem [{\citenamefont {Ezawa}(2025)}]{pwave-PhysRevB.111.L161301}%
  \BibitemOpen
  \bibfield  {author} {\bibinfo {author} {\bibfnamefont {M.}~\bibnamefont {Ezawa}},\ }\bibfield  {title} {\bibinfo {title} {Out-of-plane edelstein effects: Electric field induced magnetization in $p$-wave magnets},\ }\href {https://doi.org/10.1103/PhysRevB.111.L161301} {\bibfield  {journal} {\bibinfo  {journal} {Phys. Rev. B}\ }\textbf {\bibinfo {volume} {111}},\ \bibinfo {pages} {L161301} (\bibinfo {year} {2025})}\BibitemShut {NoStop}%
\bibitem [{\citenamefont {Ma}\ \emph {et~al.}(2026)\citenamefont {Ma}, \citenamefont {Li}, \citenamefont {Xu}, \citenamefont {Zhang},\ and\ \citenamefont {Jia}}]{maHyperspinAltermagnets2026}%
  \BibitemOpen
  \bibfield  {author} {\bibinfo {author} {\bibfnamefont {H.-Y.}\ \bibnamefont {Ma}}, \bibinfo {author} {\bibfnamefont {Y.}~\bibnamefont {Li}}, \bibinfo {author} {\bibfnamefont {H.}~\bibnamefont {Xu}}, \bibinfo {author} {\bibfnamefont {S.}~\bibnamefont {Zhang}},\ and\ \bibinfo {author} {\bibfnamefont {J.-F.}\ \bibnamefont {Jia}},\ }\bibfield  {title} {\bibinfo {title} {Hyperspin {{Altermagnets}}},\ }\href {https://doi.org/10.1103/dnr2-gv31} {\bibfield  {journal} {\bibinfo  {journal} {Phys. Rev. Lett.}\ }\textbf {\bibinfo {volume} {137}},\ \bibinfo {pages} {116704} (\bibinfo {year} {2026})}\BibitemShut {NoStop}%
\bibitem [{\citenamefont {Novoselov}\ \emph {et~al.}(2005)\citenamefont {Novoselov}, \citenamefont {Geim}, \citenamefont {Morozov}, \citenamefont {Jiang}, \citenamefont {Katsnelson}, \citenamefont {Grigorieva}, \citenamefont {Dubonos},\ and\ \citenamefont {Firsov}}]{Novoselov2005Nature}%
  \BibitemOpen
  \bibfield  {author} {\bibinfo {author} {\bibfnamefont {K.~S.}\ \bibnamefont {Novoselov}}, \bibinfo {author} {\bibfnamefont {A.~K.}\ \bibnamefont {Geim}}, \bibinfo {author} {\bibfnamefont {S.~V.}\ \bibnamefont {Morozov}}, \bibinfo {author} {\bibfnamefont {D.}~\bibnamefont {Jiang}}, \bibinfo {author} {\bibfnamefont {M.~I.}\ \bibnamefont {Katsnelson}}, \bibinfo {author} {\bibfnamefont {I.~V.}\ \bibnamefont {Grigorieva}}, \bibinfo {author} {\bibfnamefont {S.~V.}\ \bibnamefont {Dubonos}},\ and\ \bibinfo {author} {\bibfnamefont {A.~A.}\ \bibnamefont {Firsov}},\ }\bibfield  {title} {\bibinfo {title} {Two-dimensional gas of massless {Dirac} fermions in graphene},\ }\href {https://doi.org/10.1038/nature04233} {\bibfield  {journal} {\bibinfo  {journal} {Nature}\ }\textbf {\bibinfo {volume} {438}},\ \bibinfo {pages} {197} (\bibinfo {year} {2005})}\BibitemShut {NoStop}%
\bibitem [{\citenamefont {Lu}\ \emph {et~al.}(2024)\citenamefont {Lu}, \citenamefont {Reddy}, \citenamefont {Jeon}, \citenamefont {Mazza}, \citenamefont {Brahlek}, \citenamefont {Wu}, \citenamefont {Yang}, \citenamefont {Cook}, \citenamefont {Conner}, \citenamefont {Zhang}, \citenamefont {Chakraborty}, \citenamefont {Yao}, \citenamefont {Tien}, \citenamefont {Tseng}, \citenamefont {Yang}, \citenamefont {Lien}, \citenamefont {Lin}, \citenamefont {Chiang}, \citenamefont {Vignale}, \citenamefont {Li}, \citenamefont {Chang}, \citenamefont {Moore},\ and\ \citenamefont {Bian}}]{luRealizationTwodimensionalWeyl2024a}%
  \BibitemOpen
  \bibfield  {author} {\bibinfo {author} {\bibfnamefont {Q.}~\bibnamefont {Lu}}, \bibinfo {author} {\bibfnamefont {P.~V.~S.}\ \bibnamefont {Reddy}}, \bibinfo {author} {\bibfnamefont {H.}~\bibnamefont {Jeon}}, \bibinfo {author} {\bibfnamefont {A.~R.}\ \bibnamefont {Mazza}}, \bibinfo {author} {\bibfnamefont {M.}~\bibnamefont {Brahlek}}, \bibinfo {author} {\bibfnamefont {W.}~\bibnamefont {Wu}}, \bibinfo {author} {\bibfnamefont {S.~A.}\ \bibnamefont {Yang}}, \bibinfo {author} {\bibfnamefont {J.}~\bibnamefont {Cook}}, \bibinfo {author} {\bibfnamefont {C.}~\bibnamefont {Conner}}, \bibinfo {author} {\bibfnamefont {X.}~\bibnamefont {Zhang}}, \bibinfo {author} {\bibfnamefont {A.}~\bibnamefont {Chakraborty}}, \bibinfo {author} {\bibfnamefont {Y.-T.}\ \bibnamefont {Yao}}, \bibinfo {author} {\bibfnamefont {H.-J.}\ \bibnamefont {Tien}}, \bibinfo {author} {\bibfnamefont {C.-H.}\ \bibnamefont {Tseng}}, \bibinfo {author} {\bibfnamefont {P.-Y.}\ \bibnamefont {Yang}}, \bibinfo {author} {\bibfnamefont {S.-W.}\ \bibnamefont
  {Lien}}, \bibinfo {author} {\bibfnamefont {H.}~\bibnamefont {Lin}}, \bibinfo {author} {\bibfnamefont {T.-C.}\ \bibnamefont {Chiang}}, \bibinfo {author} {\bibfnamefont {G.}~\bibnamefont {Vignale}}, \bibinfo {author} {\bibfnamefont {A.-P.}\ \bibnamefont {Li}}, \bibinfo {author} {\bibfnamefont {T.-R.}\ \bibnamefont {Chang}}, \bibinfo {author} {\bibfnamefont {R.~G.}\ \bibnamefont {Moore}},\ and\ \bibinfo {author} {\bibfnamefont {G.}~\bibnamefont {Bian}},\ }\bibfield  {title} {\bibinfo {title} {Realization of a two-dimensional {{Weyl}} semimetal and topological {{Fermi}} strings},\ }\href {https://doi.org/10.1038/s41467-024-50329-6} {\bibfield  {journal} {\bibinfo  {journal} {Nat. Commun.}\ }\textbf {\bibinfo {volume} {15}},\ \bibinfo {pages} {6001} (\bibinfo {year} {2024})}\BibitemShut {NoStop}%
\bibitem [{\citenamefont {Gao}\ \emph {et~al.}(2021)\citenamefont {Gao}, \citenamefont {Wu}, \citenamefont {Persson},\ and\ \citenamefont {Wang}}]{gaoIrvspObtainIrreducible2021}%
  \BibitemOpen
  \bibfield  {author} {\bibinfo {author} {\bibfnamefont {J.}~\bibnamefont {Gao}}, \bibinfo {author} {\bibfnamefont {Q.}~\bibnamefont {Wu}}, \bibinfo {author} {\bibfnamefont {C.}~\bibnamefont {Persson}},\ and\ \bibinfo {author} {\bibfnamefont {Z.}~\bibnamefont {Wang}},\ }\bibfield  {title} {\bibinfo {title} {Irvsp: {{To}} obtain irreducible representations of electronic states in the {{VASP}}},\ }\href {https://doi.org/10.1016/j.cpc.2020.107760} {\bibfield  {journal} {\bibinfo  {journal} {Comput. Phys. Comm.}\ }\textbf {\bibinfo {volume} {261}},\ \bibinfo {pages} {107760} (\bibinfo {year} {2021})}\BibitemShut {NoStop}%
\bibitem [{\citenamefont {Zhang}\ \emph {et~al.}(2023)\citenamefont {Zhang}, \citenamefont {Sheng}, \citenamefont {Song}, \citenamefont {Liang}, \citenamefont {Jiang}, \citenamefont {Sun}, \citenamefont {Wu}, \citenamefont {Weng}, \citenamefont {Fang}, \citenamefont {Dai},\ and\ \citenamefont {Wang}}]{zhangZhangVASP2KPKpModels2023}%
  \BibitemOpen
  \bibfield  {author} {\bibinfo {author} {\bibfnamefont {S.}~\bibnamefont {Zhang}}, \bibinfo {author} {\bibfnamefont {H.}~\bibnamefont {Sheng}}, \bibinfo {author} {\bibfnamefont {Z.-D.}\ \bibnamefont {Song}}, \bibinfo {author} {\bibfnamefont {C.}~\bibnamefont {Liang}}, \bibinfo {author} {\bibfnamefont {Y.}~\bibnamefont {Jiang}}, \bibinfo {author} {\bibfnamefont {S.}~\bibnamefont {Sun}}, \bibinfo {author} {\bibfnamefont {Q.}~\bibnamefont {Wu}}, \bibinfo {author} {\bibfnamefont {H.}~\bibnamefont {Weng}}, \bibinfo {author} {\bibfnamefont {Z.}~\bibnamefont {Fang}}, \bibinfo {author} {\bibfnamefont {X.}~\bibnamefont {Dai}},\ and\ \bibinfo {author} {\bibfnamefont {Z.}~\bibnamefont {Wang}},\ }\bibfield  {title} {\bibinfo {title} {{{VASP2KP}}: $k\cdot p$ {{Models}} and {{Landé}} g-{{Factors}} from ab initio {{Calculations}}},\ }\href {https://doi.org/10.1088/0256-307X/40/12/127101} {\bibfield  {journal} {\bibinfo  {journal} {Chin. Phys. Lett.}\ }\textbf {\bibinfo {volume} {40}},\ \bibinfo {pages} {127101} (\bibinfo
  {year} {2023})}\BibitemShut {NoStop}%
\bibitem [{\citenamefont {Zhao}\ \emph {et~al.}(2025)\citenamefont {Zhao}, \citenamefont {Li}, \citenamefont {Li}, \citenamefont {Dai}, \citenamefont {Dong}, \citenamefont {Zhou}, \citenamefont {Wang}, \citenamefont {Zhang}, \citenamefont {Ji}, \citenamefont {Zhang}, \citenamefont {Yu}, \citenamefont {Wei}, \citenamefont {Li}, \citenamefont {Li}, \citenamefont {Huang}, \citenamefont {Wang}, \citenamefont {Liu}, \citenamefont {Chen}, \citenamefont {Zhang}, \citenamefont {Wang}, \citenamefont {Li}, \citenamefont {Yang}, \citenamefont {Shi}, \citenamefont {Pan}, \citenamefont {Du}, \citenamefont {Du},\ and\ \citenamefont {Zhang}}]{zhaoRealization2DMetals2025a}%
  \BibitemOpen
  \bibfield  {author} {\bibinfo {author} {\bibfnamefont {J.}~\bibnamefont {Zhao}}, \bibinfo {author} {\bibfnamefont {L.}~\bibnamefont {Li}}, \bibinfo {author} {\bibfnamefont {P.}~\bibnamefont {Li}}, \bibinfo {author} {\bibfnamefont {L.}~\bibnamefont {Dai}}, \bibinfo {author} {\bibfnamefont {J.}~\bibnamefont {Dong}}, \bibinfo {author} {\bibfnamefont {L.}~\bibnamefont {Zhou}}, \bibinfo {author} {\bibfnamefont {Y.}~\bibnamefont {Wang}}, \bibinfo {author} {\bibfnamefont {P.}~\bibnamefont {Zhang}}, \bibinfo {author} {\bibfnamefont {K.}~\bibnamefont {Ji}}, \bibinfo {author} {\bibfnamefont {Y.}~\bibnamefont {Zhang}}, \bibinfo {author} {\bibfnamefont {H.}~\bibnamefont {Yu}}, \bibinfo {author} {\bibfnamefont {Z.}~\bibnamefont {Wei}}, \bibinfo {author} {\bibfnamefont {J.}~\bibnamefont {Li}}, \bibinfo {author} {\bibfnamefont {X.}~\bibnamefont {Li}}, \bibinfo {author} {\bibfnamefont {Z.}~\bibnamefont {Huang}}, \bibinfo {author} {\bibfnamefont {B.}~\bibnamefont {Wang}}, \bibinfo {author} {\bibfnamefont {J.}~\bibnamefont
  {Liu}}, \bibinfo {author} {\bibfnamefont {Y.}~\bibnamefont {Chen}}, \bibinfo {author} {\bibfnamefont {X.}~\bibnamefont {Zhang}}, \bibinfo {author} {\bibfnamefont {S.}~\bibnamefont {Wang}}, \bibinfo {author} {\bibfnamefont {N.}~\bibnamefont {Li}}, \bibinfo {author} {\bibfnamefont {W.}~\bibnamefont {Yang}}, \bibinfo {author} {\bibfnamefont {D.}~\bibnamefont {Shi}}, \bibinfo {author} {\bibfnamefont {J.}~\bibnamefont {Pan}}, \bibinfo {author} {\bibfnamefont {S.}~\bibnamefont {Du}}, \bibinfo {author} {\bibfnamefont {L.}~\bibnamefont {Du}},\ and\ \bibinfo {author} {\bibfnamefont {G.}~\bibnamefont {Zhang}},\ }\bibfield  {title} {\bibinfo {title} {Realization of {{2D}} metals at the Ångström thickness limit},\ }\href {https://doi.org/10.1038/s41586-025-08711-x} {\bibfield  {journal} {\bibinfo  {journal} {Nature}\ }\textbf {\bibinfo {volume} {639}},\ \bibinfo {pages} {354} (\bibinfo {year} {2025})}\BibitemShut {NoStop}%
\bibitem [{\citenamefont {Wang}\ \emph {et~al.}(2016)\citenamefont {Wang}, \citenamefont {Alexandradinata}, \citenamefont {Cava},\ and\ \citenamefont {Bernevig}}]{wangHourglassFermions2016}%
  \BibitemOpen
  \bibfield  {author} {\bibinfo {author} {\bibfnamefont {Z.}~\bibnamefont {Wang}}, \bibinfo {author} {\bibfnamefont {A.}~\bibnamefont {Alexandradinata}}, \bibinfo {author} {\bibfnamefont {R.~J.}\ \bibnamefont {Cava}},\ and\ \bibinfo {author} {\bibfnamefont {B.~A.}\ \bibnamefont {Bernevig}},\ }\bibfield  {title} {\bibinfo {title} {Hourglass fermions},\ }\href {https://doi.org/10.1038/nature17410} {\bibfield  {journal} {\bibinfo  {journal} {Nature}\ }\textbf {\bibinfo {volume} {532}},\ \bibinfo {pages} {189} (\bibinfo {year} {2016})}\BibitemShut {NoStop}%
\bibitem [{\citenamefont {Zhou}\ \emph {et~al.}(2020)\citenamefont {Zhou}, \citenamefont {Zhang},\ and\ \citenamefont {Law}}]{PhysRevApplied.13.024053}%
  \BibitemOpen
  \bibfield  {author} {\bibinfo {author} {\bibfnamefont {B.~T.}\ \bibnamefont {Zhou}}, \bibinfo {author} {\bibfnamefont {C.-P.}\ \bibnamefont {Zhang}},\ and\ \bibinfo {author} {\bibfnamefont {K.}~\bibnamefont {Law}},\ }\bibfield  {title} {\bibinfo {title} {Highly tunable nonlinear hall effects induced by spin-orbit couplings in strained polar transition-metal dichalcogenides},\ }\href {https://doi.org/10.1103/PhysRevApplied.13.024053} {\bibfield  {journal} {\bibinfo  {journal} {Phys. Rev. Appl.}\ }\textbf {\bibinfo {volume} {13}},\ \bibinfo {pages} {024053} (\bibinfo {year} {2020})}\BibitemShut {NoStop}%
\bibitem [{\citenamefont {Fu}\ \emph {et~al.}(2023)\citenamefont {Fu}, \citenamefont {Cong}, \citenamefont {Xu}, \citenamefont {Zhu}, \citenamefont {Zhao}, \citenamefont {Liu}, \citenamefont {Yao}, \citenamefont {Xu}, \citenamefont {Deng}, \citenamefont {Zhu}, \citenamefont {Wang}, \citenamefont {Kang}, \citenamefont {Zeng}, \citenamefont {Lin}, \citenamefont {Wang}, \citenamefont {Tang}, \citenamefont {Yang}, \citenamefont {Dong}, \citenamefont {Liu}, \citenamefont {Xiong}, \citenamefont {Zhou}, \citenamefont {Wang}, \citenamefont {Li}, \citenamefont {Tan}, \citenamefont {Tay},\ and\ \citenamefont {Liu}}]{fuBerryCurvatureDipole2023}%
  \BibitemOpen
  \bibfield  {author} {\bibinfo {author} {\bibfnamefont {Q.}~\bibnamefont {Fu}}, \bibinfo {author} {\bibfnamefont {X.}~\bibnamefont {Cong}}, \bibinfo {author} {\bibfnamefont {X.}~\bibnamefont {Xu}}, \bibinfo {author} {\bibfnamefont {S.}~\bibnamefont {Zhu}}, \bibinfo {author} {\bibfnamefont {X.}~\bibnamefont {Zhao}}, \bibinfo {author} {\bibfnamefont {S.}~\bibnamefont {Liu}}, \bibinfo {author} {\bibfnamefont {B.}~\bibnamefont {Yao}}, \bibinfo {author} {\bibfnamefont {M.}~\bibnamefont {Xu}}, \bibinfo {author} {\bibfnamefont {Y.}~\bibnamefont {Deng}}, \bibinfo {author} {\bibfnamefont {C.}~\bibnamefont {Zhu}}, \bibinfo {author} {\bibfnamefont {X.}~\bibnamefont {Wang}}, \bibinfo {author} {\bibfnamefont {L.}~\bibnamefont {Kang}}, \bibinfo {author} {\bibfnamefont {Q.}~\bibnamefont {Zeng}}, \bibinfo {author} {\bibfnamefont {M.}~\bibnamefont {Lin}}, \bibinfo {author} {\bibfnamefont {X.}~\bibnamefont {Wang}}, \bibinfo {author} {\bibfnamefont {B.}~\bibnamefont {Tang}}, \bibinfo {author} {\bibfnamefont {J.}~\bibnamefont
  {Yang}}, \bibinfo {author} {\bibfnamefont {Z.}~\bibnamefont {Dong}}, \bibinfo {author} {\bibfnamefont {F.}~\bibnamefont {Liu}}, \bibinfo {author} {\bibfnamefont {Q.}~\bibnamefont {Xiong}}, \bibinfo {author} {\bibfnamefont {J.}~\bibnamefont {Zhou}}, \bibinfo {author} {\bibfnamefont {Q.}~\bibnamefont {Wang}}, \bibinfo {author} {\bibfnamefont {X.}~\bibnamefont {Li}}, \bibinfo {author} {\bibfnamefont {P.}~\bibnamefont {Tan}}, \bibinfo {author} {\bibfnamefont {B.~K.}\ \bibnamefont {Tay}},\ and\ \bibinfo {author} {\bibfnamefont {Z.}~\bibnamefont {Liu}},\ }\bibfield  {title} {\bibinfo {title} {Berry {{Curvature Dipole Induced Giant Mid}}‐{{Infrared Second}}‐{{Harmonic Generation}} in {{2D Weyl Semiconductor}}},\ }\href {https://doi.org/10.1002/adma.202306330} {\bibfield  {journal} {\bibinfo  {journal} {Adv. Mater.}\ }\textbf {\bibinfo {volume} {35}},\ \bibinfo {pages} {2306330} (\bibinfo {year} {2023})}\BibitemShut {NoStop}%
\bibitem [{\citenamefont {Jin}\ \emph {et~al.}(2021)\citenamefont {Jin}, \citenamefont {Oh}, \citenamefont {Stania}, \citenamefont {Liu},\ and\ \citenamefont {Yeom}}]{jinEnhancedBerryCurvature2021}%
  \BibitemOpen
  \bibfield  {author} {\bibinfo {author} {\bibfnamefont {K.-H.}\ \bibnamefont {Jin}}, \bibinfo {author} {\bibfnamefont {E.}~\bibnamefont {Oh}}, \bibinfo {author} {\bibfnamefont {R.}~\bibnamefont {Stania}}, \bibinfo {author} {\bibfnamefont {F.}~\bibnamefont {Liu}},\ and\ \bibinfo {author} {\bibfnamefont {H.~W.}\ \bibnamefont {Yeom}},\ }\bibfield  {title} {\bibinfo {title} {Enhanced {{Berry Curvature Dipole}} and {{Persistent Spin Texture}} in the {{Bi}}(110) {{Monolayer}}},\ }\href {https://doi.org/10.1021/acs.nanolett.1c02811} {\bibfield  {journal} {\bibinfo  {journal} {Nano Lett.}\ }\textbf {\bibinfo {volume} {21}},\ \bibinfo {pages} {9468} (\bibinfo {year} {2021})}\BibitemShut {NoStop}%
\bibitem [{\citenamefont {Kim}\ \emph {et~al.}(2019)\citenamefont {Kim}, \citenamefont {Kim}, \citenamefont {Shin}, \citenamefont {Lee}, \citenamefont {Sinova}, \citenamefont {Park},\ and\ \citenamefont {Jin}}]{kimPredictionFerroelectricitydrivenBerry2019d}%
  \BibitemOpen
  \bibfield  {author} {\bibinfo {author} {\bibfnamefont {J.}~\bibnamefont {Kim}}, \bibinfo {author} {\bibfnamefont {K.-W.}\ \bibnamefont {Kim}}, \bibinfo {author} {\bibfnamefont {D.}~\bibnamefont {Shin}}, \bibinfo {author} {\bibfnamefont {S.-H.}\ \bibnamefont {Lee}}, \bibinfo {author} {\bibfnamefont {J.}~\bibnamefont {Sinova}}, \bibinfo {author} {\bibfnamefont {N.}~\bibnamefont {Park}},\ and\ \bibinfo {author} {\bibfnamefont {H.}~\bibnamefont {Jin}},\ }\bibfield  {title} {\bibinfo {title} {Prediction of ferroelectricity-driven {{Berry}} curvature enabling charge- and spin-controllable photocurrent in tin telluride monolayers},\ }\href {https://doi.org/10.1038/s41467-019-11964-6} {\bibfield  {journal} {\bibinfo  {journal} {Nat Commun}\ }\textbf {\bibinfo {volume} {10}},\ \bibinfo {pages} {3965} (\bibinfo {year} {2019})}\BibitemShut {NoStop}%
\bibitem [{\citenamefont {Lai}\ \emph {et~al.}(2024)\citenamefont {Lai}, \citenamefont {Zhan}, \citenamefont {Liu}, \citenamefont {Chen},\ and\ \citenamefont {Sun}}]{laiElectricFieldTunable2024a}%
  \BibitemOpen
  \bibfield  {author} {\bibinfo {author} {\bibfnamefont {J.}~\bibnamefont {Lai}}, \bibinfo {author} {\bibfnamefont {J.}~\bibnamefont {Zhan}}, \bibinfo {author} {\bibfnamefont {P.}~\bibnamefont {Liu}}, \bibinfo {author} {\bibfnamefont {X.-Q.}\ \bibnamefont {Chen}},\ and\ \bibinfo {author} {\bibfnamefont {Y.}~\bibnamefont {Sun}},\ }\bibfield  {title} {\bibinfo {title} {Electric field tunable nonlinear {{Hall}} terahertz detector in the dual quantum spin {{Hall}} insulator {{TaIrTe}} 4},\ }\href {https://doi.org/10.1103/PhysRevB.110.155122} {\bibfield  {journal} {\bibinfo  {journal} {Phys. Rev. B}\ }\textbf {\bibinfo {volume} {110}},\ \bibinfo {pages} {155122} (\bibinfo {year} {2024})}\BibitemShut {NoStop}%
\bibitem [{\citenamefont {Sharma}\ \emph {et~al.}(2025)\citenamefont {Sharma}, \citenamefont {Ghimire}, \citenamefont {Shahi}, \citenamefont {Bhattarai}, \citenamefont {Sun}, \citenamefont {You}, \citenamefont {Van Den~Brink},\ and\ \citenamefont {Singh}}]{sharmaLargeBerryCurvature2025}%
  \BibitemOpen
  \bibfield  {author} {\bibinfo {author} {\bibfnamefont {B.}~\bibnamefont {Sharma}}, \bibinfo {author} {\bibfnamefont {M.~P.}\ \bibnamefont {Ghimire}}, \bibinfo {author} {\bibfnamefont {D.~B.}\ \bibnamefont {Shahi}}, \bibinfo {author} {\bibfnamefont {D.}~\bibnamefont {Bhattarai}}, \bibinfo {author} {\bibfnamefont {Y.}~\bibnamefont {Sun}}, \bibinfo {author} {\bibfnamefont {J.-S.}\ \bibnamefont {You}}, \bibinfo {author} {\bibfnamefont {J.}~\bibnamefont {Van Den~Brink}},\ and\ \bibinfo {author} {\bibfnamefont {S.}~\bibnamefont {Singh}},\ }\bibfield  {title} {\bibinfo {title} {Large {{Berry}} curvature dipole and enhanced nonlinear {{Hall}} effects in monolayer {{W}} 0.5 {{Mo}} 0.5 {{Te}} 2},\ }\href {https://doi.org/10.1103/99gy-krxw} {\bibfield  {journal} {\bibinfo  {journal} {Phys. Rev. B}\ }\textbf {\bibinfo {volume} {112}},\ \bibinfo {pages} {235433} (\bibinfo {year} {2025})}\BibitemShut {NoStop}%
\bibitem [{\citenamefont {Zhang}\ and\ \citenamefont {Fu}(2021)}]{zhangTerahertzDetectionBased2021}%
  \BibitemOpen
  \bibfield  {author} {\bibinfo {author} {\bibfnamefont {Y.}~\bibnamefont {Zhang}}\ and\ \bibinfo {author} {\bibfnamefont {L.}~\bibnamefont {Fu}},\ }\bibfield  {title} {\bibinfo {title} {Terahertz detection based on nonlinear {{Hall}} effect without magnetic field},\ }\href {https://doi.org/10.1073/pnas.2100736118} {\bibfield  {journal} {\bibinfo  {journal} {Proc. Natl. Acad. Sci. U.S.A.}\ }\textbf {\bibinfo {volume} {118}},\ \bibinfo {pages} {e2100736118} (\bibinfo {year} {2021})}\BibitemShut {NoStop}%
\bibitem [{\citenamefont {Kang}\ \emph {et~al.}(2025)\citenamefont {Kang}, \citenamefont {Lee},\ and\ \citenamefont {Park}}]{Kang2025-app}%
  \BibitemOpen
  \bibfield  {author} {\bibinfo {author} {\bibfnamefont {M.-G.}\ \bibnamefont {Kang}}, \bibinfo {author} {\bibfnamefont {S.}~\bibnamefont {Lee}},\ and\ \bibinfo {author} {\bibfnamefont {B.-G.}\ \bibnamefont {Park}},\ }\bibfield  {title} {\bibinfo {title} {Field-free spin-orbit torques switching and its applications},\ }\href {https://doi.org/10.1038/s44306-025-00071-6} {\bibfield  {journal} {\bibinfo  {journal} {npj Spintronics}\ }\textbf {\bibinfo {volume} {3}},\ \bibinfo {pages} {8} (\bibinfo {year} {2025})}\BibitemShut {NoStop}%
\end{thebibliography}
%


\end{document}